\documentclass[fleqn,usenatbib]{mnras}

\usepackage{newtxtext,newtxmath}

\usepackage[T1]{fontenc}

\DeclareRobustCommand{\VAN}[3]{#2}
\let\VANthebibliography\thebibliography
\def\thebibliography{\DeclareRobustCommand{\VAN}[3]{##3}\VANthebibliography}

\usepackage{graphicx}	
\usepackage{amsmath}	

\graphicspath{ {Images/} }

\def\HII{H{\sc ii} }
\defcitealias{Menon_20}{MFK2020}
\usepackage{booktabs}
\usepackage{caption}
\usepackage{threeparttable}

\title[Turbulence Driving Parameter in Carina]{On the compressive nature of turbulence driven by ionising feedback in the pillars of the Carina Nebula}

\author[S.~H.~Menon et al.]{
Shyam H.~Menon$^{1}$\thanks{E-mail: shyam.menon@anu.edu.au},
Christoph Federrath$^{1}$,
Pamela Klaassen$^{2}$,
Rolf Kuiper$^{3}$,
Megan Reiter$^{2}$
\\
$^{1}$Research School of Astronomy and Astrophysics, Australian National University, Canberra, ACT~2611, Australia\\
$^{2}$UK Astronomy Technology Centre, Royal Observatory Edinburg, Blackford Hill, Edinburgh EH9 3HJ, UK\\
$^{3}$Institute of Astronomy and Astrophysics, University of T{\"u}bingen, Auf der Morgenstelle 10, D-72076 T{\"u}bingen, Germany
}

\date{Accepted XXX. Received YYY; in original form ZZZ}

\pubyear{2020}

\begin{document}
\label{firstpage}
\pagerange{\pageref{firstpage}--\pageref{lastpage}}
\maketitle

\begin{abstract}
The ionizing radiation of massive stars sculpts the surrounding neutral gas into pillar-like structures. Direct signatures of star formation through outflows and jets are observed in these structures, typically at their tips. Recent numerical simulations have suggested that this star formation could potentially be triggered by photoionising radiation, driving compressive modes of turbulence in the pillars. In this study we use recent high-resolution ALMA observations of $^{12}\mathrm{CO}$, $^{13}\mathrm{CO}$, and $\mathrm{C}^{18} \mathrm{O}, \; J = 2-1$ emission to test this hypothesis for pillars in the Carina Nebula. We analyse column density and intensity-weighted velocity maps, and subtract any large-scale bulk motions in the plane of the sky to isolate the turbulent motions. We then reconstruct the dominant turbulence driving mode in the pillars, by computing the turbulence driving parameter $b$, characterised by the relation $\sigma_{\rho/\rho_0} = b \mathcal{M}$ between the standard deviation of the density contrast $\sigma_{\rho/\rho_0}$ (with gas density $\rho$ and its average $\rho_0$) and the turbulent Mach number $\mathcal{M}$. We find values of \mbox{$b\sim0.7$--$1.0$} for most of the pillars, suggesting that predominantly compressive modes of turbulence are driven in the pillars by the ionising radiation from nearby massive stars. We find that this range of $b$ values can produce star formation rates in the pillars that are a factor $\sim 3$ greater than with $b \sim 0.5$, a typical average value of $b$ for spiral-arm molecular clouds. Our results provide further evidence for the potential triggering of star formation in pillars through compressive turbulent motions. 
\end{abstract}




\section{Introduction}

Expanding \HII regions are known to sculpt the surrounding neutral gas in the ISM into structures reminiscent of the iconic 'Pillars of Creation' imaged by the Hubble Space Telescope \citep{Hester_1996}. There have been a wealth of observations that image these pillars and related structures such as globules, energetic evaporating globules (EEG's) and proplyds, to deduce dynamical quantities in and around them \citep{Preibisch_2012,Mann_2014,Grenman_2014,Klaassen_2014,Hartigan_2015,Schneider_2016,Djupvik_2017,Klaassen_2018,Reiter_2019,Klaassen_2020}. Models proposed to explain their formation fall broadly into two categories: the classic collect-and-collapse model by \cite{Elmegreen_1977}, where the \HII region sweeps up and accumulates cold gas, creating density enhancements and eventually pillars in their shadows; or the more recent radiation-driven implosion (RDI) model, where clouds with pre-existing density enhancements are sculpted to form pillars by impinging ionizing radiation. In these studies the density enhancements are modelled as Bonnor-Ebert spheres or are seeded by turbulence, and they naturally produce the observed morphologies and dynamics of pillars \citep{Mellema_2006,Gritschneder_2010,Mackey_2010,Walch_2012,Tremblin_2013,Menon_20}. 

Direct signatures of star formation are observed at the tip of these pillars through jets and outflows with suggestions that this star formation could be triggered by the ionizing radiation \citep{Sugitani_2002,Billot_2010,Smith_2010,Chauhan_2011,Reiter_2013,Roccatagliata_2013,Cortes-Rangel_2019}. However, one could argue whether star formation is really 'triggered', as stars could have formed by the direct gravitational collapse of the pre-existing density enhancements anyway, and the ionizing radiation need not have enhanced local star formation in any way. Indeed, numerical simulations of giant molecular cloud (GMC) evolution show that while there are some signs that photoionization feedback could trigger star formation \emph{locally}, they are spatially intermixed with spontaneously formed stars, and hence it is difficult to differentiate the two \citep{Dale_2012,Dale_2013}. Moreover, \citet{Dale_2015_Trigger} argues that current observational markers used to infer triggering may not be robust enough to distinguish whether an observed star was triggered or has spontaneously formed. Thus, there is no consensus to what extent triggered star formation is effective on GMC scales, and hence it is important to investigate the local dynamical factors that influence star formation in the pillars. 

The dominant mode of turbulence that is driven in the pillars, i.e., solenoidal (divergence-free) or compressive (curl-free), or a mixture of the two extremes, quantified by the turbulence driving parameter $b$ is one such relavant dynamical factor. The value of $b$ typically varies between 1/3 and 1, where these two extreme cases refer to purely solenoidal and purely compressive driving, respectively \citep{Federrath_2010}, and depends on the physical process driving the turbulence; for instance gravity, galactic shear, stellar feedback, etc.~\citep[see][for a review of potential drivers]{Elmegreen_2009,Federrath_2017,Federrath_2018}. The value of $b$ is important, as the flow dynamics, density structure and the subsequent star formation rate depend on it \citep{Federrath_2008,Federrath_2010,Price_2011,Konstandin_2012,Padoan_2014,Federrath_2015,Nolan_2015}; with compressive driving resulting in broader density probability distribution functions (PDFs) and star formation rates approximately an order of magnitude larger than for solenoidal driving \citep{federrath_klessen_2012,Kainulainen_2014,Federrath_2018}. In fact, the value of $b$ is often used as a proxy to explain the observed star formation rate of a cloud with higher values of $b$ corresponding to more efficient star formation and vice versa \citep[see for instance][for the case of the CMZ 'Brick' cloud's inefficient star formation]{Federrath_2016}. In this context, \citet{Menon_20} (hereafter referred to as \citetalias{Menon_20}) deduced through numerical simulations that expanding \HII regions drive predominantly compressive modes of turbulence in pillar-like structures raising the value of $b$ from its original value, and through this mechanism, could potentially trigger star formation within them. In this study, we intend to observationally test this claim by computing the value of $b$ for pillars in the Carina Nebula, as has been done for different clouds in the Milky Way \citep{Padoan_1997,Brunt_2010_Taurus,Ginsburg_2013,Kainulainen_2013,Federrath_2016,Kainulainen_2017}.

The Carina Nebula (NGC 3372) is a perfect laboratory to study the dynamical effects of feedback from massive stars, as it contains more than 65 O-type stars \citep{Smith_2006} and about $10^5 \mathrm{M}_{\sun}$ of gas and dust surrounding it that is being sculpted by the UV radiation from the stars \citep{Preibisch_2011c,Preibisch_2011b}. The majority of the stellar UV radiation in Carina originates in the massive star clusters Trumpler 14 and 16, with significant contributions from very massive stars ($M \ga 100 \, M_{\sun}$) such as the well-known $\eta$~Carinae. Carina also has copious amounts of active low-to-intermediate mass star formation occurring in the gas surrounding these massive stars, where this star formation could have been triggered by the ionising radiation \citep{Gaczkowski_2013,Roccatagliata_2013}. Most importantly, its relative proximity allows us to resolve sufficiently well the kinematic and density structure in the pillars. In a recent study, \citet{Klaassen_2020} presented the first ALMA Compact Array (ACA) survey of 13 pillars in Carina in $^{12}\mathrm{CO}$, $^{13}\mathrm{CO}$, and $\mathrm{C}^{18}\mathrm{O}, \; J =2-1$ emission that could sufficiently well resolve the detailed kinematics and density structures within the pillar. The pillars observed cover a wide range in relative position within the nebula, proximity to irradiating clusters, and morphologies.  

In this study we pick out the six best resolved candidate pillars among the 13 presented in \citet{Klaassen_2020}, and investigate the dominant mode of turbulence in them. We quantify this by calculating the turbulence driving parameter $b$ calculated from its standard relation given by \citep{Padoan_1997,Federrath_2008,Federrath_2010,Padoan_2011,Price_2011,Konstandin_2012,Molina_2012,Hopkins_2013,Federrath_2015,Nolan_2015,Squire_2017,Mandal_2019},
\begin{equation}
    \sigma_{\rho/\rho_0} = b \mathcal{M},
\end{equation}
where $\sigma_{\rho/\rho_0}$ is the standard deviation of the 3D volume density contrast $\rho/\rho_0$, and $\mathcal{M}$ is the turbulent RMS Mach number. We prepare column density and velocity maps from the ALMA position-position-velocity (PPV) data cubes, and compute the dispersion in their respective PDFs, from which the value of $b$ is calculated.  

The paper is organised as follows. In Section~\ref{sec:observational_methods} we summarise the details of the observations used in this work. Section~\ref{sec:b_method} delineates the general method we follow to quantify the density and velocity distributions in the pillars and subsequently calculate the turbulence driving parameter $b$. Section~\ref{sec:Results} presents our results for the different pillar regions. We then discuss in Section~\ref{sec:Discussion} the implications of our results to star formation in the pillars, followed by some caveats of this work. We finally summarise and conclude our results in Section~\ref{sec:Summary}.


\section{Observational data and methods}
\label{sec:observational_methods}
\subsection{ALMA Observations}
\label{sec:alma} 
The observational data for the regions used in this work are presented in detail in \citet{Klaassen_2020}. Here we briefly summarise the data. This work samples the 'Southern Pillars', 'Northern Pillars' and 'Southwestern Pillar' complexes as is defined in \citet{Hartigan_2015}, as part of the ALMA Compact Array (ACA) survey. The ACA survey observes 13 pillars in Carina in $^{12}\mathrm{CO}$, $^{13}\mathrm{CO}$, and $\mathrm{C}^{18}\mathrm{O}$, $J = 2-1$ line emission, and the 230~$\mathrm{GHz}$ continuum. \citet{Klaassen_2020} obtained both ACA and Total Power (TP) data sets which were feathered together. This method improved the sensitivity to the large-scale structure of the CO emission in the pillars. 

\subsection{Temperature maps}
\label{sec:Temperature_Maps}
In order to measure the turbulence driving parameter $b$, we need a measurement of the Mach number $\mathcal{M}=\sigma_v/c_\mathrm{s}$, i.e., the ratio of velocity dispersion and sound speed. Since the sound speed depends primarily on the gas temperature ($T_\mathrm{gas}$), we need an estimate of $T_\mathrm{gas}$. To trace $T_\mathrm{gas}$, we use the temperature maps of the large-scale dust emission in the Carina nebula from \citet{Marsh_2017}. This study maps the temperature structure of the dust throughout the galactic plane with \textit{Herschel} Hi-GAL data using the novel PPMAP procedure \citep{Marsh_2015}. The complete set of maps they produced, including the one we adopt, is available online on the VIALACTEA database \footnote{\url{http://www.astro.cardiff.ac.uk/research/ViaLactea/}}. Note that the temperature maps we show represent the mean dust temperature along the line-of-sight.

\begin{figure*}
    \centering
    \includegraphics[width=\textwidth]{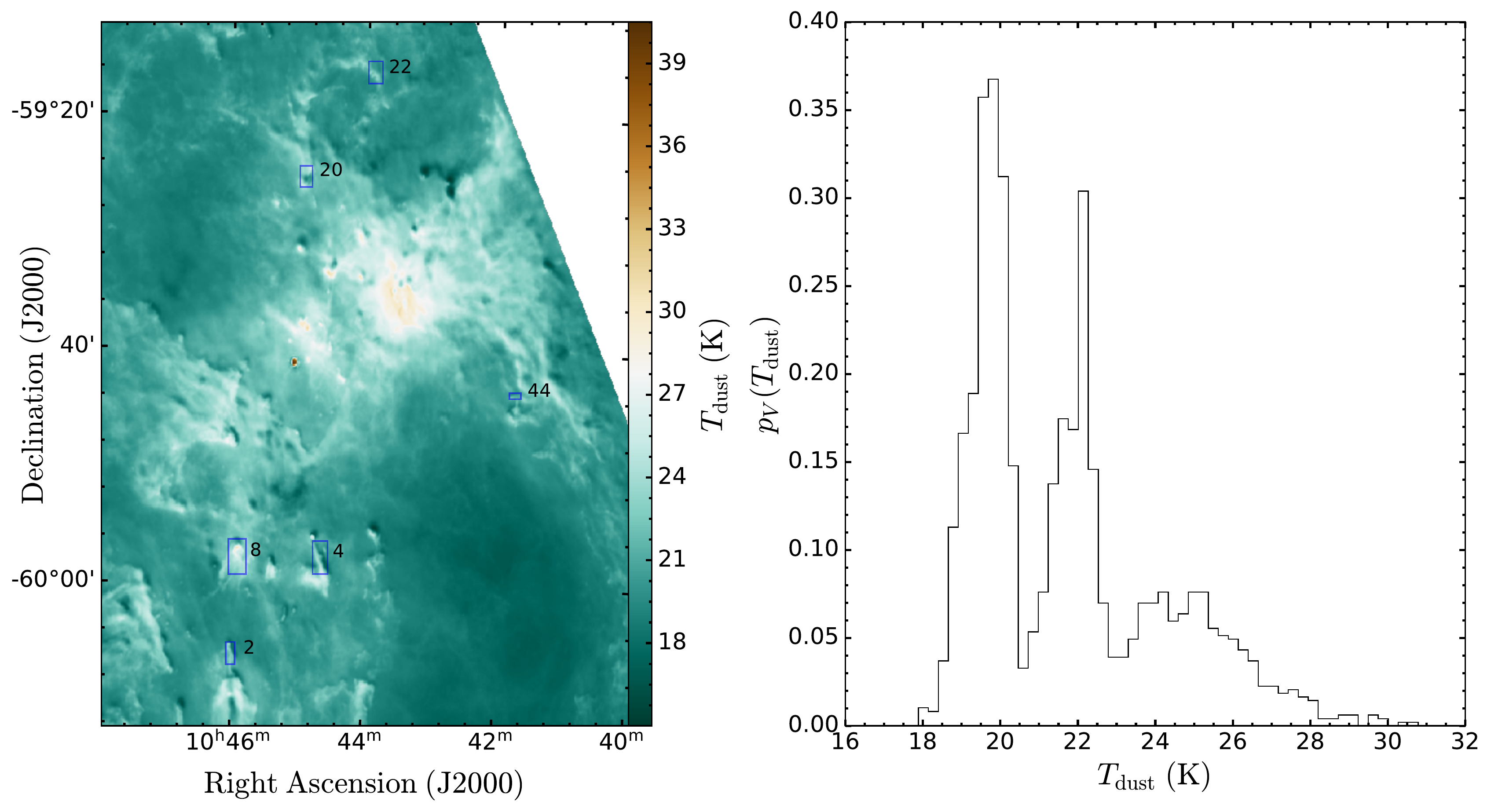}
    \caption{Large-scale temperature structure ($T_\mathrm{dust}$) of Carina obtained from \textit{Herschel} dust emission \citep{Marsh_2017} (left panel). We annotate boxes around the pillar regions that we study here, with the pillar numbers adopted from \citet{Hartigan_2015}. The PDF on the right panel shows the distribution $T_\mathrm{dust}$ from all pixels within the annotated pillar regions.}
    \label{fig:Dust_Temperature}
\end{figure*}

In Figure~\ref{fig:Dust_Temperature} we show the large-scale dust temperature distribution in a region within Carina and annotate boxes on the pillars that are studied in this paper. We also show the distribution of dust temperatures for the regions (or pixels) that lie within the boundaries of our pillar regions. We can see that the dust temperature lies roughly in the range \mbox{$T_\mathrm{dust} \sim 15$--$30 \, \mathrm{K}$}. We assume that the gas and dust must be sufficiently well coupled at the densities corresponding to the pillars, and thus use the same distribution of $T_\mathrm{dust}$ for the gas temperature, i.e., we assume $T_\mathrm{gas}=T_\mathrm{dust}$. We also checked that our results are not overly dependent on the assumed distribution for $T_\mathrm{gas}$, by comparing our results to a simpler case, where the gas temperature is assumed to be a uniform distribution in the range \mbox{$10$--$30 \, \mathrm{K}$}, instead of the measured dust distribution, and found that the median values of $b$ vary by less than 5\%, which is well within the $1\sigma$ uncertainties of our results. 

\section{Measuring the turbulence driving parameter}
\label{sec:b_method}
The driving parameter $b$ is a quantity that is proportional to the ratio of density to velocity fluctuations, $b \propto \sigma_{\rho}/\sigma_{v}$ in a supersonically turbulent cloud \citep{Federrath_2008,Federrath_2010}. A velocity field that contains primarily compressible modes would produce stronger compressions and rarefactions, and thus result in a higher spread in the density PDF than a primarily solenoidal velocity field. The driving parameter $b$ is given by the ratio between the 3D density dispersion $\sigma_{\rho/\rho_0}$ and the turbulent sonic Mach number $ \mathcal{M}$, as
\begin{equation}
	b =\frac{\sigma_{\rho/\rho_0}}{\mathcal{M}}.
	\label{eq:b_parameter}
\end{equation}
Below we describe the general method we adopt to derive the column density and velocity structure of molecular gas in the pillars, and calculate their respective density and velocity dispersions. We then discuss our approach to reconstruct $\sigma_{\rho/\rho_0}$ from a two-dimensional (2D) column density dispersion. We compute $b$ from the derived physical parameters, and discuss the uncertainties propagated to the final value of $b$. 

\subsection{Gas density structure}
\label{sec:b_method_density_structure}
To trace the $\mathrm{H}_2$ gas column density structure in the pillars, we use a combination of the $^{13}\mathrm{CO}$ and $\mathrm{C}^{18}\mathrm{O}, \, J=2-1$ line emission under the assumption that they are optically thin in their region of use. The $^{12}\mathrm{CO}$ emission is not used to directly trace the $\mathrm{H}_2$ column density in this work, primarily because $^{12}\mathrm{CO}$ is optically thick within all the pillars, and thus cannot accurately trace the true column density of $\mathrm{H}_2$ \citep[however see][for an alternative method with a correction factor for optically-thick lines]{Mangum2015}. In addition, we are interested in the turbulent dynamics of the dense, potentially star-forming gas, whereas the $^{12}\mathrm{CO}$ emission would include significant contributions from the more diffuse emission towards the periphery of the pillars.

The choice of whether to use $^{13}\mathrm{CO}$ or $\mathrm{C}^{18}\mathrm{O}$ in each spectral channel and spatial region is dependent on the $^{13}\mathrm{CO}$ optical thickness ($\tau_{13}$). The optical depth of $^{13}\mathrm{CO}$ can be estimated as $\tau_{13} = \tau_{12}X[^{13}\mathrm{CO}/^{12}\mathrm{CO}]$, where $\tau_{12}$ is the optical depth of $^{12}\mathrm{CO}$, which we calculate using equation~1 in \citet{Choi_1993}, and $X[^{13}\mathrm{CO}/^{12}\mathrm{CO}]$ is the abundance ratio of $^{13}\mathrm{CO}$ to $^{12}\mathrm{CO}$. Thus, we can determine whether $^{13}\mathrm{CO}$ is optically thick (i.e., $\tau_{13}>1$) in a region, if $\tau_{12}$ is greater than the abundance ratio $X[^{12}\mathrm{CO}/^{13}\mathrm{CO}] = 1/X[^{13}\mathrm{CO}/^{12}\mathrm{CO}]$, in which case we use $\mathrm{C}^{18}\mathrm{O}$ to trace the $\mathrm{H}_2$ column density in that region. Otherwise, we use $^{13}\mathrm{CO}$. To convert the respective isotopologue column density to an $\mathrm{H}_2$ column density, we compute the abundance multiplication factor $X [\mathrm{H}_2]$ based on the line chosen to trace $\mathrm{H}_\mathrm{2}$, and is given as
\begin{equation}
    X [\mathrm{H}_2] = 
    \begin{cases}
        X [^{12}\mathrm{CO}/^{13}\mathrm{CO}] \times X[\mathrm{H}_2/^{12}\mathrm{CO}] & \text{if} \;\; \tau_{13} \leq 1 \\
        X [^{12}\mathrm{CO}/\mathrm{C} ^{18} \mathrm{O}] \times X[\mathrm{H}_2/^{12}\mathrm{CO}] & \text{if} \;\; \tau_{13} > 1
    \end{cases} .
\end{equation}
We use abundances $X [^{12}\mathrm{CO}/^{13}\mathrm{CO}] = 60$ \citep{Rebolledo_2015}, $X [^{12}\mathrm{CO}/\mathrm{C}^{18}\mathrm{O}] = 560$ \citep{Wilson_1994}, and $X[\mathrm{H}_2/^{12}\mathrm{CO}] = 1.1 \times 10^4$ \citep{Pineda2010}.

We apply a signal-to-noise cutoff corresponding to $3 \sigma_{\mathrm{noise}}$ for all our maps, where $\sigma_{\mathrm{noise}} \sim 2.1 \, \mathrm{Jy}\,\mathrm{beam}^{-1}$ is the noise in the $^{13}\mathrm{CO}$ data cube, and corresponds to an equivalent column density threshold $N > 1.1 \times 10^{20}\, \mathrm{cm}^{-2}$. Once the noise is filtered out, the relevant isotopologue among $^{13}\mathrm{CO}$ and $\mathrm{C}^{18} \mathrm{O}$ in each spectral channel and spatial region is chosen based on the optical thickness conditions described above. We then compute the $\mathrm{H}_2$ column density from the main-beam brightness temperature $T_\mathrm{MB}$ of the relevant line expressed in units of Kelvin ($\mathrm{K}$) as
\begin{equation}
\label{eq:Column_Density}
N_{\mathrm{H}_2} = X [\mathrm{H}_2] \int f(T_\mathrm{ex})  T_\mathrm{MB} \, dv ,
\end{equation}
where the pre-factor $f(T_\mathrm{ex})$ is a function of the excitation temperature ($T_{\mathrm{ex}}$). The function $f$ is given by \citet{Mangum2015} as
\begin{equation}
\label{eq:Prefactor}
	f \left(T_{\mathrm{ex}} \right) = \frac{3hZ}{8 \pi^3 \mu^2 J_u} \frac{\exp\left(\frac{E_\mathrm{up}}{kT_{\mathrm{ex}}}\right) }{1-\exp \left( \frac{h\nu}{kT_{\mathrm{ex}}} \right)} \frac{1}{J \left(T_{\mathrm{ex}} \right)-J\left(T_\mathrm{BG} \right)} ,
\end{equation}
where $\nu$ is the rest-frame line frequency, $\mu$ the molecule dipole moment, $Z$ the rotational partition function, which can be approximated as $Z = kT_{\mathrm{ex}}/hB + 1/3$ where $B = \nu/2J_u$ is the rotational constant, $E_\mathrm{up}$ the energy of the upper rotational level, $J_u = 2$ the upper quantum level number, and $J(T_{\mathrm{ex}})-J(T_\mathrm{BG})$ the correction for background CMB radiation, where
\begin{equation}
\label{eq:J_equation}
	J(T) = \frac{h\nu}{k \left(\exp\left(\frac{h\nu}{kT}\right)-1 \right)} .
\end{equation}
The values we use for the above quantities in the calculation are given in Appendix \ref{sec:appendix_line_constants}. For the excitation temperature we assume LTE and use a conservative uniform distribution range of \mbox{$T_{\mathrm{ex}} \sim 10$ -- $30\,\mathrm{K}$}, motivated by the range of dust temperatures we observe in the pillars (Figure~\ref{fig:Dust_Temperature}). 

Once the column density maps are computed, the mean $N_0$ is computed for each of them, and is used to compute $\sigma_\eta$, the dispersion of the natural logarithm of the column density scaled by its mean, i.e., $\eta = \ln \left( N/N_0 \right)$. The dispersion $\sigma_\eta$ is estimated by fitting a \citet{Hopkins_2013} intermittency density PDF model to the volume-weighted PDF of $\eta$. This is a physically motivated fitting function that takes into account intermittent features in the density PDF, and provides a more accurate description of the density structure compared to the lognormal approximation, even at high turbulent Mach numbers and equations of state that are not isothermal \citep{Hopkins_2013,Federrath_2015}. The Hopkins fitting function is given by
\begin{multline}
\label{eq:Hopkins}
    p_{\mathrm{HK}}(\eta)\,d\eta = I_1 \left(2 \sqrt{\lambda \omega(\eta)}\right) \exp {\left[ -\left(\lambda + \omega(\eta)\right) \right]} \sqrt{\frac{\lambda}{\theta^2 \omega(\eta)}}\,d\eta, \\
    \lambda = \frac{\sigma_{\eta}^2}{2 \theta^2}, \quad \omega(\eta) = \lambda/(1+\theta) - \eta/\theta \;\; (\omega \geq 0),
\end{multline}
where $I_1(x)$ is the first-order modified Bessel Function of the first kind, $\sigma_{\eta}$ is the standard deviation in $\eta$, and $\theta$ is the intermittency parameter that encapsulates the intermittent density PDF features. Note that in the zero-intermittency limit ($\theta \to 0$), Eq.~(\ref{eq:Hopkins}) simplifies to the lognormal PDF. We compute the best-fit parameters of $\sigma_\eta$ and $\theta$, and use them to transform $\sigma_\eta$ to the linear dispersion $\sigma_{N/N_0}$ by using the relation \citep{Hopkins_2013},
\begin{equation}
\label{eq:sigma_N_eta_relation}
    \sigma_{N/N_0} = \sqrt{\exp \left( {\frac{\sigma_\eta^2}{1+3\theta +2 \theta^2}} \right) -1} ,
\end{equation}
which can be derived from the moments of Eq.(~\ref{eq:Hopkins}). 

Alternatively, studies have also shown that the column density PDF of molecular gas in GMCs can be traced by a piecewise form consisting of a lognormal followed by a power law at high densities that accounts for self-gravitating gas \citep{Klessen_2000,Federrath_2008_Tracer,Kritsuk_2011,Federrath_Klessen_2013,Girichidis_2014,Burkhart_2016}. We attempted to fit our distributions of $\eta$ for the pillars with this functional form, but found that the fit was not accurate, and a Hopkins density PDF performed significantly better. This could potentially be due to the limitations of using $\mathrm{CO}$ and its isotopologues to trace molecular gas at high densities, as $\mathrm{CO}$ is prone to optical depth effects and/or depletion at higher column densities \citep[see for instance,][]{Schneider2016}. We also point out there are additional caveats in using interferometric data to measure column density PDFs \citep[e.g.,][]{Ossenkopf_2016}, however correcting for them are beyond the scope of this work.

\subsection{Conversion from two-dimensional to three-dimensional density dispersion}
\label{sec:b_method_brunts_method}
We do not have direct access to the 3D (volume) density from observations, but rather only to the 2D (column) density distribution. However, it is possible to estimate the 3D density dispersion and the density PDF from the 2D column density information given in the plane of the sky to the third dimension (along the line of sight), by assuming isotropy of the clouds \citep{Brunt_2010a,Brunt_2010b,Kainulainen_2014}. Here we use the \citet{Brunt_2010b} method for these purposes.

To estimate the 3D density dispersion ($\sigma_{\rho/\rho_0}$) from the 2D column density dispersion ($\sigma_{N/N_0}$), we first measure the 2D column density power spectrum $P_\mathrm{2D}(k)$ of the variable $N/N_0 - 1$, where $k$ is the wavenumber. This is used to reconstruct the 3D density power spectrum $P_\mathrm{3D}(k)$ of the variable $\rho/\rho_0 -1$, as $P_\mathrm{3D}(k) = 2kP_\mathrm{2D}(k)$. The variance of a mean-zero field is proportional to the integral of the power spectrum over all $k$ \citep{Arfken_2013}. Thus, the ratio $\mathcal{R}^{1/2}$ of the 3D density dispersion ($\sigma_{\rho/\rho_0}$) and the 2D column density dispersion ($\sigma_{N/N_0}$) is defined as
\begin{equation}
\label{eq:Brunt_Equation}
    \mathcal{R}^{1/2} = \frac{\sigma_{N/N_0}}{\sigma_{\rho/\rho_0}} = \frac{\sum_k P_\mathrm{2D}(k)}{\sum_k P_\mathrm{3D}(k)} .
\end{equation}
\citet{Brunt_2010b} showed that Eq.(~\ref{eq:Brunt_Equation}) holds to within 10\% for isotropic, periodic fields, and is less accurate for non-periodic fields. We thus apply mirroring of the column density field to generate a periodic dataset \citep{Ossenkopf_2008}. For each pillar, we also check the robustness of the assumption of isotropy in the column density field. To estimate the degree of anisotropy in the density structure, we fit elliptical contours to the 2D Fourier image and measure the aspect ratio from the ratio of their major and minor axes, to estimate the degree of uncertainty as a measure of length scale. We find a range of axis ratios in our pillars, including some with values $\sim 1.5$ that correspond to anistropic structures. Indeed, numerical simulations have demonstrated that strong magnetic guide fields can produce aspect ratios up to $\sim 2.0$, in which case the maximum uncertainty in the 2D-to-3D reconstruction for these cases is $<40\%$ \citep{Federrath_2016}. We thus conservatively adopt this upper limit as a systematic uncertainty in the 2D-to-3D reconstruction in this study. This reconstructed $\sigma_{\rho/\rho_0}$ is used in conjunction with an independent turbulent Mach number $\mathcal{M}$ calculation discussed in the following.

\subsection{Gas Kinematics}
\label{sec:b_method_kinematics}
To estimate the turbulent velocity dispersion $\sigma_v$ of the molecular gas, we use the first moment map, i.e., the intensity-weighted line-of-sight (LOS) velocity map of the molecular gas in the pillars traced with a combination of the $^{13}\mathrm{CO}$ and $\mathrm{C}^{18}\mathrm{O}$ lines based on their optical depths as discussed in Section~\ref{sec:b_method_density_structure}. Unlike the more conventional method, we do not use the mean value of the second moment map as the estimate of the LOS $\sigma_v$. This is because the LOS velocities traced by the second moment map could have contributions from bulk motions on the scale of the pillars due to shear, systematic rotation, gravitational infall, outflows, etc. These bulk motions need to be subtracted to obtain the true turbulent velocity dispersion ($\sigma_v$) inside the pillar . This is because these motions encompass the scale of the whole pillar, and thus are not appropriate for studying the local turbulent dynamics in the pillars, set by an external physical mechanism acting on or above the scales of the pillar itself. In this sense, the first-moment map helps in recognising and fitting bulk-motion velocity profiles (if any) in the plane-of-sky, that could be subtracted to obtain the purely turbulent gas motions along the LOS \citep[the same method has been applied in][]{Federrath_2016,Sharda_2018}. We estimate a first-order bulk velocity profile (i.e., a linear functional form) by fitting a plane to the values of the first-moment map for each pillar. We then subtract this fitted profile from the first-moment map to isolate the turbulent velocities. Note that while this is an improvement over the conventional method, there could be higher-order contributions from non-turbulent motions, and in a sense we could still be underestimating the systematic motions, and thus overestimating the turbulent $\sigma_v$. 

Once the turbulent motions are isolated, we should get a reasonable Gaussian profile in the LOS velocity PDF as predicted for turbulence. We estimate the turbulent LOS velocity dispersion $\sigma_{v,1\mathrm{D}}$ by fitting a Gaussian function to this PDF, and convert this 1D velocity dispersion to a 3D turbulent velocity dispersion $\sigma_{v,3\mathrm{D}} = 3^{1/2}\sigma_v$, again assuming isotropy, similar to the volume-density reconstruction discussed above. Concerning the assumption of isotropy, we checked the level of anisotropy in the column density distributions, by using Fourier analysis such as that shown in Figure 3 of \citet{Federrath_2016}, and find that the density field is reasonably isotropic for the pillar regions studied here \citep[see also,][]{Brunt_2010b}. Thus, we believe that in the absence of further information, we have to assume that we can also treat the turbulent velocity field as being reasonably isotropic. The computed $\sigma_{v,3\mathrm{D}}$ is then divided by the thermal sound speed $c_\mathrm{s}$ to obtain the turbulent RMS Mach number $\mathcal{M}$. As discussed in Section~\ref{sec:Temperature_Maps} we adopt a gas temperature of $T_\mathrm{gas}=T_\mathrm{dust}$ where $T_\mathrm{dust}$ has the distribution shown in Figure~\ref{fig:Dust_Temperature}. This corresponds to a non-uniform distribution for the thermal sound speed in the range \mbox{$c_\mathrm{s} \sim 0.18$ -- $0.30 \, \mathrm{km}/\mathrm{s}$}, assuming a mean molecular weight of $2.8 \, m_{\mathrm{H}}$; an appropriate value for a cloud of 71\% molecular hydrogen gas, 27\% helium and 2\% metal abundance \citep[e.g.,][]{Kauffmann_2008}.

Note that Gaussian fits to the turbulent velocity PDFs may have some residual deviation. These deviations from a purely Gaussian PDF may be due to intrinsic noise and uncertainties, varying excitation conditions for the line emissions used to trace the molecular gas, inaccuracies in the adopted bulk motion profile, and intrinsic 'intermittency' in the turbulence, especially in the tails of the distributions \citep[see for instance][]{Passot_1998,Kritsuk_2007,Federrath_2009,Federrath_2010,Federrath_2013,Hopkins_2013}.

\subsection{Propagating uncertainties to the calculation of $b$}
\label{sec:b_method_uncertainties}

To propagate the various sources of systematic uncertainties involved in the calculation of $b$, we use a fully Monte-Carlo based error propagation scheme. In this scheme, we essentially model all directly measured quantities/parameters as Gaussian priors, with a mean corresponding to the true value of the quantity, and standard deviation corresponding to the RMS uncertainty in the same quantity. We then draw Monte-Carlo samples from this Gaussian prior for the parameter/quantity used to derive another parameter/quantity. This procedure generates random realisations of the new quantity, obeying the underlying relationship between the generating quantities and the new quantity. The resulting realisations are propagated across the entire calculation pipeline of $b$, as opposed to computing the errors in each step by assuming that they are Gaussian-distributed around the mean as in standard error propagation. We report the value of each derived quantity as the median of its realisations, and the statistical errors as $1-\sigma$ equivalent percentiles (i.e., $16^{\mathrm{th}}$ and $84^{\mathrm{th}}$ percentiles). The advantage of this is that propagating the realisations using Monte-Carlo sampling maintains any inherent skewness or non-Gaussian features in the error residual distribution of a derived quantity. We use 10,000 Monte-Carlo sampled realisations for propagating the errors, which we found was sufficient for convergence. We discuss some inherent uncertainties that we propagate in this manner, relevant to the calculation of $b$. 

The most straightforward source of uncertainty to quantify is the flux calibration errors of the data set. \citet{Klaassen_2020} report that while the uncertainties varied slightly between observing dates and calibrators, the upper limit was less than 6\%. We use this upper envelope in this work. We also propagate the uncertainty due to the intrinsic discrete nature of the spectral channels, by assuming a uniform distribution prior for the spectral line frequency/velocity, whose width corresponds to the spectral channel width ($\sim 0.06 \, \mathrm{km}/\mathrm{s}$). Similarly, as discussed earlier, the gas temperature ($T_\mathrm{gas}$) is represented by the dust temperature distribution $T_\mathrm{dust}$, and the excitation temperature ($T_\mathrm{ex}$) is sampled from a uniform distribution in the range \mbox{ $10$ -- $30\,  \mathrm{K}$}. For computing the mass in the pillar we use a distance to Carina of $d = 2.5 \pm 0.28 \, \mathrm{kpc}$, as estimated in \citet{Povich2019} using recent \textit{Gaia} measurements. Further, errors estimated from the covariance matrix in least-squares fitting of Gaussian/Hopkins functions to our PDFs are combined in quadrature with the inherent error in the respective quantity.

\section{Results}
\label{sec:Results}



In this section, we discuss the results we obtain from the general analysis method discussed in Section~\ref{sec:b_method} for each individual pillar region. The numbering scheme of the pillars is adopted from \citet{Hartigan_2015}, and is identical to the one used in \citet{Klaassen_2020}. For each pillar we first discuss general derived physical parameters such as the effective area ($A$), derived mass $M$, 3D line-of-sight (LOS) velocity dispersion obtained along the LOS $\sigma_{v,3\mathrm{D}}^{\mathrm{mom}2}$ from the second-moment map, LOS velocity dispersion across the plane-of-sky $\sigma_{v,3\mathrm{D}}$ from the first-moment map as described in Section~\ref{sec:b_method_kinematics}, the corresponding virial parameter using this velocity dispersion $\alpha_{\mathrm{vir}}$, and the ratio of gas free-fall time $t_\mathrm{ff}$ and the turbulent crossing time $t_{\mathrm{turb}}$. We then proceed to derive the physical quantities relevant to the calculation of the turbulence driving parameter ($b$), as outlined in Section~\ref{sec:b_method}. The summary of the physical quantities, values of $b$ and the parameters involved in its calculation for all the pillars are given in Table~\ref{tab:pillars_summary}. Note that the quantities quoted in the table and the conclusions we derive from them are only for pixels that fall within the $1.1 \times 10^{22} \, \mathrm{cm}^{-2}$ (3$\sigma$ sensitivity) column density cutoff contour we use to filter noise (see Section~\ref{sec:b_method_density_structure}). However, we show in Appendix~\ref{sec:appendix_SNcutoff} that our conclusions still hold irrespective of the particular choice of this threshold. 

\subsection{Pillar 20}
\label{sec:Pillar20}

Pillar~20 is a relatively large and thus well-resolved pillar observed in the study of \citet{Klaassen_2020} that lies spatially in the "Northern Pillars" region as characterised in \citet{Hartigan_2015}. It forms a part of the dust rim surrounding the massive star cluster Trumpler~15. We derive an effective area of $\sim 0.63 \, \mathrm{pc}^2$, a mean column density of $N_0 \sim 10^{22} \, \mathrm{cm}^{-2}$, and a corresponding effective mass of $M \sim 140 \, M_{\sun}$. There are already potential signs of ongoing star formation in this pillar. For instance, \citet{Ohlendorf_2012} identified a compact green object (CGO) due to excess emission in the \textit{Spitzer} 4.5~$\micron$ band at the tip of this pillar, which could possibly be tracing shock-excited $\mathrm{H}_{2}$ of a protostellar jet driven by an intermediate-mass protostar (which they estimate to lie in the range \mbox{$\sim 4.6$--$7.7 \mathrm{M}_{\sun}$}). In addition, there is an eastward spur in the pillar that shows signs of red-shifted emission that could potentially be indicative of a protostellar outflow \citep{Klaassen_2020}. 

\begin{figure*}
    \centering
    \includegraphics[width=0.88 \textwidth]{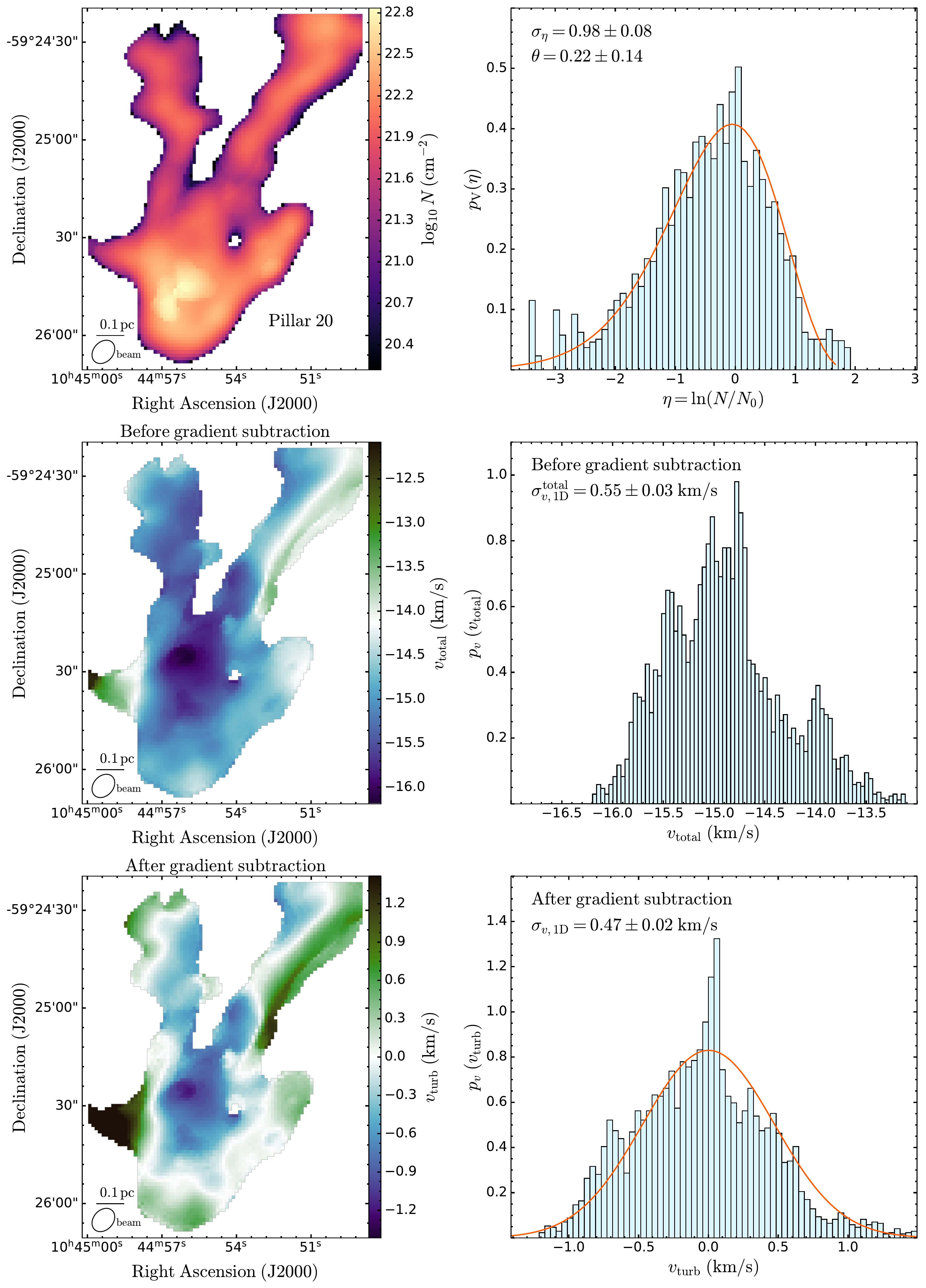}
    \caption{Pillar 20: Top panels show the $\mathrm{H}_2$ column density ($N$) map (left), and the PDF of $\eta = \ln (N/N_0)$, i.e., the logarithmic column density scaled by its mean $N_0$ (right). The corresponding Hopkins density PDF (Equation~\ref{eq:Hopkins}) fit to $\eta$ is overplotted with the best-fit parameters annotated on the plot. Middle panels show the intensity-weighted velocity (first-moment map) before subtraction of systematic motions approximated by a large-scale gradient (left), its corresponding PDF and the calculated velocity dispersion (without fitting). Bottom panels show the same after subtraction of the large-scale gradient, with the best-fit Gaussian overplotted, and the corresponding 1D velocity dispersion $\sigma_{v,1\mathrm{D}}$ annotated. This particular pillar does not show a strong systematic large-scale gradient, and hence, the correction by subtraction of the fitted gradient has a relatively small effect on this pillar. For an example where the gradient-subtraction is crucial, see Fig.~\ref{fig:Pillar22_Maps}.}
    \label{fig:Pillar20_Maps}
\end{figure*}

The derived column density and intensity-weighted velocity (first-moment) maps of Pillar~20 are shown in Figure~\ref{fig:Pillar20_Maps}. While the density structure of Pillar~20 is relatively smooth, there is an evident overdensity in the south tip of the pillar that could possibly be signs of future star formation. From the column density map and its mean value $N_0$, we compute the volume-weighted PDF of $\eta = \ln(N/N_0)$, shown in the top-right panel of Figure~\ref{fig:Pillar20_Maps}. As discussed in Section~\ref{sec:b_method_density_structure}, a Hopkins density PDF function (see Equation~\ref{eq:Hopkins}) is then fitted to the PDF of $\eta$ to compute the best-fit value of $\sigma_\eta$ and the intermittency parameter $\theta$, which are used to obtain the linear scaled column density dispersion $\sigma_{N/N_0}$, following Equation~\ref{eq:sigma_N_eta_relation}. We obtain $\sigma_\eta = 0.98(+0.08/-0.01)$, $\theta = 0.22 \pm 0.14$, and $\sigma_{N/N_0} = 0.85(+0.20/-0.09)$. The best-fit Hopkins function overplotted on the PDF of $\eta$ is also shown in Figure~\ref{fig:Pillar20_Maps}.  

Similarly, the 1D LOS turbulent velocity dispersion in the plane-of-sky, $\sigma_{v,1\mathrm{D}}$, is obtained by fitting a Gaussian distribution to the volume-weighted PDF of the LOS velocity in the first-moment map. As discussed in Section~\ref{sec:b_method_kinematics}, a linear plane-of-sky bulk velocity profile is fitted to the first-moment map and then subtracted to isolate the turbulent motions. The LOS-velocity maps before and after this subtraction are shown in the middle and lower panels of Figure~\ref{fig:Pillar20_Maps}, alongside their volume-weighted PDFs and corresponding best-fit Gaussian parameters. The fitted bulk motion profile we obtained was fairly smooth without a significant gradient in the plane-of-sky, and the subtraction was primarily removing a mean bulk velocity component in the line-of-sight. Indeed, the shape of the velocity PDFs, and the values of the velocity dispersions we obtain before and after the subtraction are quite similar for this particular pillar. Our fitted 1D purely turbulent velocity dispersion $\sigma_{v,1\mathrm{D}} = 0.47(+0.01/-0.02) \, \mathrm{km}/\mathrm{s}$ corresponds to a 3D velocity dispersion of $\sigma_{v,3\mathrm{D}} = 0.82(+0.02/-0.04) \, \mathrm{km}/\mathrm{s}$, assuming isotropy in the velocity field. This is in reasonable agreement (considering the uncertainties) with that obtained from the second-moment map (i.e., dispersion along the line-of-sight) $\sigma_{v,3\mathrm{D}}^{\mathrm{mom2}} = 1.01 \pm 0.22 \, \mathrm{km}/\mathrm{s}$. However, this is only the case for this particular pillar, which does not show a prominent large-scale gradient. A counterexample follows, for pillar~22, where the gradient-subtraction is significant and necessary to isolate the turbulent motions from contamination by systematic motions, such as rotation or large-scale shear \citep[c.f., the detailed discussion of this in][]{Federrath_2016}.

With our calculated value of $\sigma_{v,3\mathrm{D}}$ and our adopted sound speed distribution, we obtain a turbulent Mach number $\mathcal{M} = \sigma_{v,3\mathrm{D}}/c_\mathrm{s} = 3.4(+0.8/-0.5) \, \mathrm{km}/\mathrm{s}$. The resulting turbulence driving parameter $b$ is calculated using Equation~\ref{eq:b_parameter} as $b = 0.9(+0.6/-0.3)$. This suggests that the turbulence in pillar~20 is primarily compressive in nature, i.e., $b>0.4$ at a very high confidence level, considering the lower ($16^\mathrm{th}$ percentile) limit.

\subsection{Pillar 22}
\label{sec:Pillar22}

\begin{figure*}
    \centering
    \includegraphics[width=0.9\textwidth]{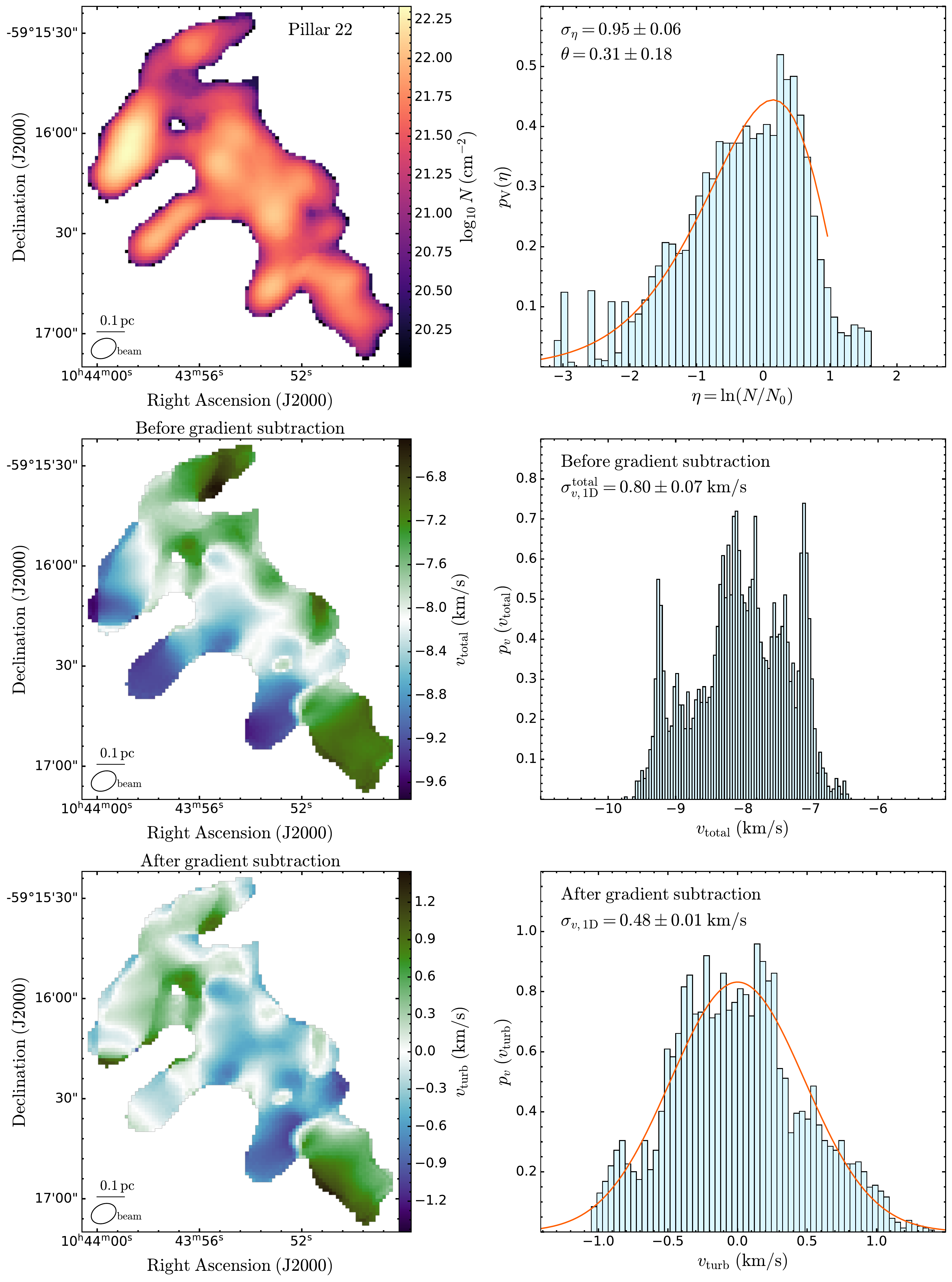}
    \caption{Same as Figure~\ref{fig:Pillar20_Maps}, but for pillar~22.}
    \label{fig:Pillar22_Maps}
\end{figure*}

Pillar~22 again lies in the "Northern Pillars" region and on the same dust-rim as pillar~20 but on the opposite side of Trumpler~15 \citep{Hartigan_2015}. It contains several finger-like protrusions that are attached to the rim of the cloud, as the structure is still being actively sculpted by the ionising radiation. The mean density and mass of the pillar is relatively low, further suggesting that the pillar is still relatively young (see Table~\ref{tab:pillars_summary}). There are no reported active sites of star formation in pillar~22, although \citet{Klaassen_2020} reports strong $\mathrm{C}^{18}\mathrm{O}$ emission in the tip of the middle protruding structure, which may potentially be signs of future star formation. The column density and first-moment maps for the pillar are shown in Figure~\ref{fig:Pillar22_Maps}. The most evident feature in this pillar is the clear velocity gradient in the direction from the protruding fingers to the base of the rim (i.e., $\sim 45 \degr$ to either axis of the image). This gradient could be the result of the fingers being pushed onto the heavier cloud rim by the thermal pressure of the photo-ionised gas \citep{Klaassen_2020}. Irrespective of the source of the motions, it is crucial that a fitted subtraction is done to obtain the relevant turbulent velocity dispersion \citep{Federrath_2016}.

\begin{figure}
    \centering
    \includegraphics[width=\columnwidth]{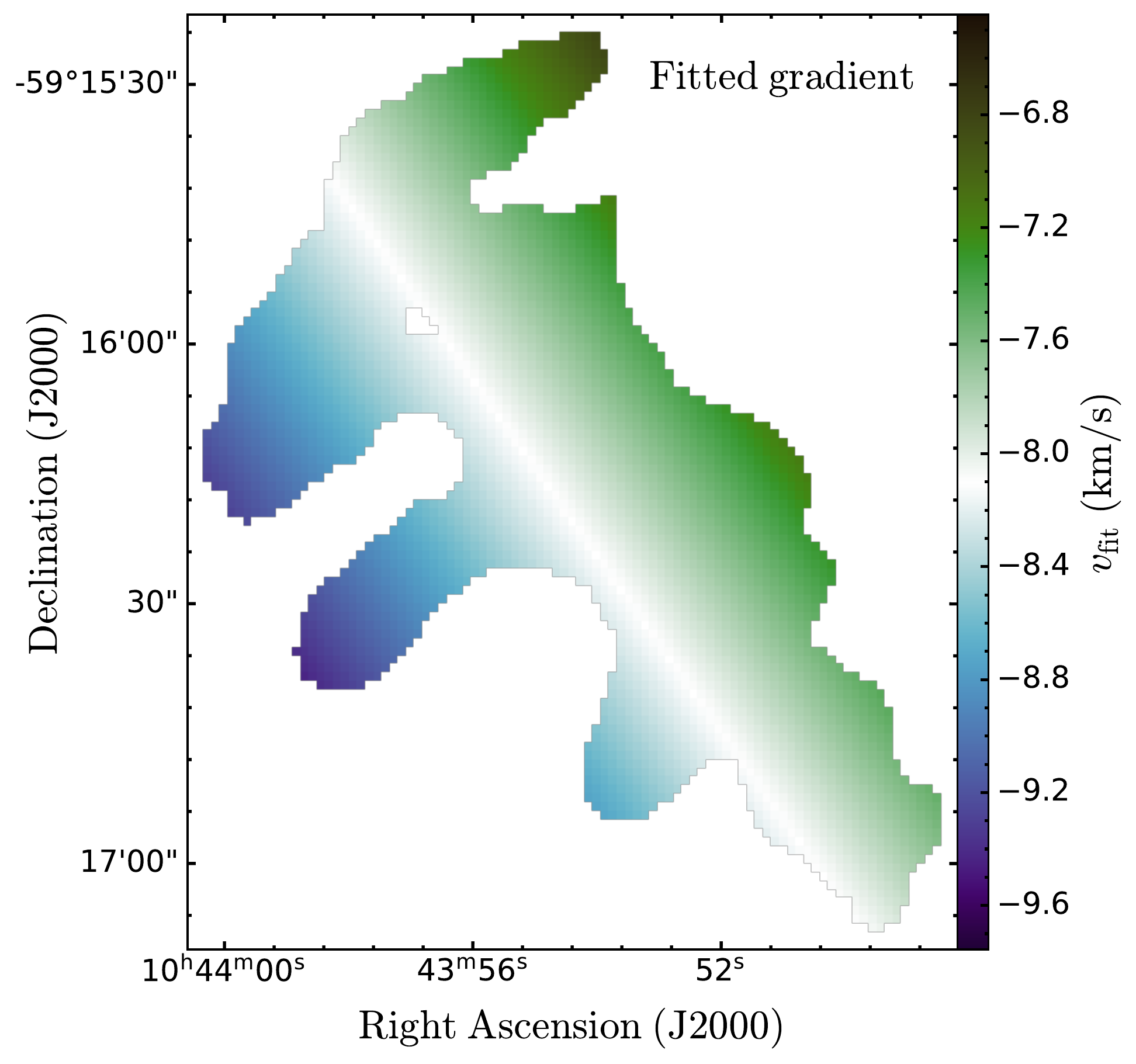}
    \caption{Fitted large-scale bulk velocity profile in the plane-of-sky $v_\mathrm{fit}$ for pillar~22. This large-scale gradient is likely caused by systematic motions such as large-scale shear or rotation, and is therefore subtracted from the first-moment map, in order to isolate the turbulent velocities.}
    \label{fig:Pillar22_Gradient}
\end{figure}

We show the PDF of $\eta$ and $v_{\mathrm{LOS}}$ (before and after subtraction) alongside the column density and first-moment maps in Figure~\ref{fig:Pillar22_Maps}. It is evident from the velocity PDF that there are significant non-turbulent contributions to the velocity structure, and thus, the total velocity dispersion is greater than the purely turbulent velocity dispersion. We show the fitted bulk-motion velocity map that is subtracted in Figure~\ref{fig:Pillar22_Gradient}, to demonstrate our method of subtraction and its importance in this application. We find that after subtraction, the velocity PDF is relatively Gaussian and we obtain a best-fit estimate of $\sigma_{v,1\mathrm{D}} = 0.48 \pm 0.01 \, \mathrm{km}\, /\mathrm{s}$. This corresponds to a $\sigma_{v,3\mathrm{D}} = 0.84 \pm 0.02 \, \mathrm{km}\, /\mathrm{s}$, and leads to a turbulent Mach number $\mathcal{M} = 3.5(+0.8/-0.5)$. With the independently derived value of $\sigma_{\rho/\rho_0} = 3.0 (+2.0/-0.9)$, we obtain a value of $b = 0.9 (+0.6/-0.3)$ for pillar~22, very similar to that for pillar~20. Pillar~22 has a virial parameter $\alpha_{\mathrm{vir}}\sim3$ compared to pillar~20 with $\alpha_{\mathrm{vir}}\sim1$ (see Table~\ref{tab:pillars_summary}), and thus, pillar~22 is more dominated by turbulent motions than pillar~20. However, we speculate that the compressive dominated turbulence in pillar~22 is currently seeding local overdensities, which could lead to star formation in the near future.

\renewcommand{\arraystretch}{1.7}
\begin{table*}
\centering
\begin{threeparttable}
\caption{Summarised comparison of physical parameters for the pillars.}
\label{tab:pillars_summary}
\begin{tabular}{l c|c c c c c c}
\toprule
\multicolumn{1}{l}{(1)}& \multicolumn{1}{c}{Pillar}& \multicolumn{1}{c}{2} & \multicolumn{1}{c}{4} & \multicolumn{1}{c}{8} & \multicolumn{1}{c}{20} & \multicolumn{1}{c}{22} & \multicolumn{1}{c}{44}\\ 
\midrule
(2) &$A \, \left[\mathrm{pc}^{2} \right]$ &$ 0.33^{+0.08}_ {-0.07} $ & $ 0.85^{+0.20}_ {-0.18} $ & $ 1.33^{+0.31}_ {-0.28} $ & $ 0.63^{+0.15}_ {-0.13} $ & $ 0.56^{+0.13}_ {-0.12} $ & $ 0.18^{+0.04}_ {-0.04} $ \\ 
(3) &$N_0 \, \left[10^{21} \mathrm{cm}^{-2} \right]$ &$ 15.5^{+3.5}_ {-3.4} $ & $ 8.6^{+1.9}_ {-1.8} $ & $ 17.5^{+3.9}_ {-3.8} $ & $ 10.0^{+2.3}_ {-2.2} $ & $ 4.4^{+1.0}_ {-1.0} $ & $ 10.1^{+2.3}_ {-2.2} $ \\ 
(4) &$M \, \left[\mathrm{M}_{\odot} \right]$ &$ 110^{+ 40}_ {- 30} $ & $ 160^{+ 60}_ {- 50} $ & $ 510^{+200}_ {-100} $ & $ 140^{+ 50}_ {- 40} $ & $  54^{+ 20}_ {- 20} $ & $  39^{+ 10}_ {- 10} $ \\ 
(5) &$n \, \left[10^{3}\mathrm{cm}^{-3} \right]$ &$ 7.8^{+2.1}_ {-1.8} $ & $ 2.7^{+0.7}_ {-0.6} $ & $ 4.4^{+1.2}_ {-1.0} $ & $ 3.6^{+1.0}_ {-0.9} $ & $ 1.7^{+0.4}_ {-0.4} $ & $ 6.9^{+1.8}_ {-1.6} $ \\ 
(6) &${\sigma_{v,\mathrm{3D}}^{\mathrm{M2}}} \, \, \left[\mathrm{km} / \mathrm{s} \right]$ &$ 1.0^{+0.3}_ {-0.3} $ & $ 1.6^{+0.5}_ {-0.5} $ & $ 1.0^{+0.2}_ {-0.2} $ & $ 1.0^{+0.2}_ {-0.2} $ & $ 0.8^{+0.1}_ {-0.1} $ & $ 0.8^{+0.1}_ {-0.1} $ \\ 
(7) &$\alpha_{\mathrm{vir}}$ &$ 0.5^{+0.2}_ {-0.1} $ & $ 1.6^{+1.0}_ {-0.4} $ & $ 0.4^{+0.1}_ {-0.1} $ & $ 1.2^{+0.4}_ {-0.3} $ & $ 3.2^{+1.0}_ {-0.7} $ & $ 0.4^{+0.1}_ {-0.1} $ \\ 
(8) &$t_{\mathrm{ff}}/t_{\mathrm{turb}}$ &$ 0.29^{+0.05}_ {-0.03} $ & $ 0.55^{+0.14}_ {-0.08} $ & $ 0.28^{+0.04}_ {-0.03} $ & $ 0.48^{+0.07}_ {-0.05} $ & $ 0.76^{+0.11}_ {-0.08} $ & $ 0.26^{+0.04}_ {-0.03} $ \\ 
(9) &${\sigma_{\eta}}$ & $ 1.17^{+0.04}_ {-0.02} $ & $ 1.05^{+0.06}_ {-0.04} $ & $ 0.81^{+0.07}_ {-0.01} $ & $ 0.98^{+0.08}_ {-0.01} $ & $ 0.95^{+0.06}_ {-0.03} $ & $ 0.80^{+0.11}_ {-0.17} $ \\ 
(10) &${\sigma_{N/N_0}}$ & $ 0.92^{+0.31}_ {-0.06} $ & $ 0.80^{+0.17}_ {-0.06} $ & $ 0.79^{+0.13}_ {-0.08} $ & $ 0.85^{+0.20}_ {-0.09} $ & $ 0.76^{+0.15}_ {-0.05} $ & $ 0.57^{+0.09}_ {-0.04} $ \\ 
(11) &${\mathcal{R}}$ & $ 0.26^{+0.11}_ {-0.10} $ & $ 0.27^{+0.11}_ {-0.11} $ & $ 0.24^{+0.10}_ {-0.09} $ & $ 0.29^{+0.12}_ {-0.11} $ & $ 0.26^{+0.10}_ {-0.10} $ & $ 0.35^{+0.14}_ {-0.14} $ \\ 
(12) &${\sigma_{\rho/\rho_0}}$ & $ 3.7^{+2.7}_ {-1.2} $ & $ 3.1^{+2.1}_ {-0.9} $ & $ 3.3^{+2.2}_ {-1.0} $ & $ 3.0^{+2.0}_ {-0.9} $ & $ 3.0^{+2.0}_ {-0.9} $ & $ 1.7^{+1.1}_ {-0.5} $ \\ 
(13) &$\sigma_{v,\mathrm{1D}}^{\mathrm{total}} \, \left[\mathrm{km} / \mathrm{s} \right]$ &$ 0.32^{+0.02}_ {-0.02} $ & $ 0.63^{+0.04}_ {-0.04} $ & $ 0.58^{+0.02}_ {-0.02} $ & $ 0.55^{+0.03}_ {-0.03} $ & $ 0.80^{+0.07}_ {-0.07} $ & $ 0.15^{+0.02}_ {-0.02} $ \\ 
(14) &$\sigma_{v,\mathrm{1D}} \, \left[\mathrm{km} / \mathrm{s} \right]$ &$ 0.30^{+0.02}_ {-0.00} $ & $ 0.51^{+0.14}_ {-0.02} $ & $ 0.43^{+0.01}_ {-0.01} $ & $ 0.47^{+0.01}_ {-0.02} $ & $ 0.48^{+0.01}_ {-0.01} $ & $ 0.19^{+0.01}_ {-0.01} $ \\ 
(15) &$\sigma_{v,\mathrm{3D}} \, \left[\mathrm{km} / \mathrm{s} \right]$ &$ 0.51^{+0.03}_ {-0.01} $ & $ 0.89^{+0.25}_ {-0.03} $ & $ 0.75^{+0.02}_ {-0.01} $ & $ 0.82^{+0.02}_ {-0.04} $ & $ 0.84^{+0.02}_ {-0.01} $ & $ 0.32^{+0.02}_ {-0.01} $ \\ 
(16) &${\mathcal{M}}$ &$ 2.2^{+0.5}_ {-0.3} $ & $ 3.9^{+1.0}_ {-0.7} $ & $ 3.1^{+0.7}_ {-0.4} $ & $ 3.4^{+0.8}_ {-0.5} $ & $ 3.5^{+0.8}_ {-0.5} $ & $ 1.3^{+0.3}_ {-0.2} $ \\ 
(17) &$b$ &$ 1.7^{+1.3}_ {-0.6} $ & $ 0.8^{+0.6}_ {-0.3} $ & $ 1.1^{+0.7}_ {-0.4} $ & $ 0.9^{+0.6}_ {-0.3} $ & $ 0.9^{+0.6}_ {-0.3} $ & $ 1.2^{+0.8}_ {-0.4} $ \\ 
\bottomrule
\end{tabular}
\begin{tablenotes}
\small
\item \textbf{Notes}: Values reported are the median along with the $16^{\mathrm{th}}$ and $84^{\mathrm{th}}$ percentiles of the distribution of realisations of each physical parameter as subscripts and superscripts, respectively.
Note that the derived physical quantities are for pixels that fall within the $1.1 \times 10^{22} \, \mathrm{cm}^{-2}$ (3$\sigma$ sensitivity) column density contour we use to filter noise.
Row (1): Pillar region number following the numbering scheme of \citet{Hartigan_2015}.
Row (2): Effective area $A = \sum l_x l_y$ where $l_x=l_y= 0.012 \, \mathrm{pc}$ are pixel sizes.
Row (3): Mean column density $N_0$ traced using a combination of $^{13}\mathrm{CO}$ and $\mathrm{C}^{18}\mathrm{O}$ emission.
Row (4): Derived mass from column density using distance to Carina $d = 2.5 \pm 0.28 \, \mathrm{kpc}$.
Row (5): Volume number density $n = N_0/L$ where $N_0$ is the mean column density and $L = 2(A/\pi)^{1/2}$ the effective diameter.
Row (6): 3D velocity dispersion in the line-of-sight obtained from the mean value of the second-moment map.
Row (7): Virial Parameter $\alpha_{\mathrm{vir}} = 5{\sigma_{v,\mathrm{3D}}}^{2}/(\pi G L^2 \rho_0)$ where $\rho_0 = n_0 \mu_{\mathrm{mol}} m_{\mathrm{H}}$ is the mass density.
Row (8): Ratio of the gas freefall time $t_{\mathrm{ff}} = {\left[ 3\pi/(32G\rho_0) \right]}^{1/2}$ and turbulent crossing time $t_{\mathrm{turb}} = L/\sigma_{v,\mathrm{3D}}$.
Row (9): Dispersion of $\eta = \ln (N/N_0)$, i.e., logarithmic column density scaled by its mean $N_0$, obtained by fitting a Hopkins function.
Row (10): $\sigma_{N/N_0}$, the dispersion of the column density scaled by its mean from $\sigma_\eta$ and $\theta$, the fitted Hopkins intermittency parameter using Eq.~(\ref{eq:sigma_N_eta_relation}).
Row (11): Brunt's factor for 2D-to-3D reconstruction of density field given by Eq.~(\ref{eq:Brunt_Equation}).
Row (12): 3D volume density dispersion $\sigma_{\rho/\rho_0} = \sigma_{N/N_0}/\mathcal{R}^{1/2}$.
Row (13): 1D velocity dispersion in the plane-of-sky obtained from the first-moment map, \textit{before} large scale correction.
Row (14): 1D turbulent velocity dispersion in the plane-of-sky obtained from the first-moment map, \textit{after} large scale correction.
Row (15): 3D turbulent velocity dispersion in the plane-of-sky obtained after large scale gradient correction as $\sigma_{v,\mathrm{3D}} = 3^{1/2}\sigma_{v,\mathrm{1D}}$.
Row (16): Turbulent RMS Mach no $\mathcal{M} = \sigma_{v,\mathrm{3D}}/c_\mathrm{s}$ where $c_\mathrm{s}$ is the thermal gas sound speed.
Row (17): Turbulence driving parameter $b = \sigma_{\rho/\rho_0}/\mathcal{M}$, where $b \sim 0.33$: purely solenoidal, $b \sim 1.0$: purely compressive and  $b \sim 0.4$ a mixture of both.
Note that values of $b>1$ suggest that there are contributions to the compressive density structure from other agents such as gravity.
\end{tablenotes}
\end{threeparttable}
\end{table*}



\subsection{Pillar 2}
\label{sec:Pillar2}

Pillar~2 is a member of the "South Pillars" as categorised in \citet{Hartigan_2015}, a region known to have a higher-than-average young stellar object (YSO) population \citep{Smith_2010_Spitzer}. This large dust pillar (G287.88-0.93) roughly points towards $\eta$~Carinae and marks the relatively narrow top of an even larger structure known as the 'Giant Pillar' \citep[G287.93-0.99; see][]{Smith_2000}, which contains the most massive concentration of molecular gas in the entire region. Similar to the other Southern Pillars, there are signs of active and ongoing star formation in pillar~2, traced through protostellar jets and dense $\mathrm{C}^{18}\mathrm{O}$ cores that are virially unstable \citep{Klaassen_2020}. Pillar~2 is host to the HH~903 jet from an embedded protostar in the middle-west edge of the pillar, with a total outflow length of $\sim 2 \, \mathrm{pc}$, one of the longest parsec-scale outflows discovered in the Carina Nebula \citep{Smith_2010}. However, the molecular gas does not show any signs of outflow motions associated with the jet, which suggests that the jet has un-coupled from the embedded material. In addition, \citet{Reiter_2017} discovered another jet, HH~1169 \citep[candidate jet HHc-10 in][]{Smith_2010} at the top-right tip of the pillar. We refer the reader to figure~5 in \citet{Reiter_2017} for a rough idea of the locations of the jets in pillar~2, and \citet{Reiter_2016} for the estimated positions of the protostars driving them.  

\begin{figure*}
    \centering
    \includegraphics[width=0.8\textwidth]{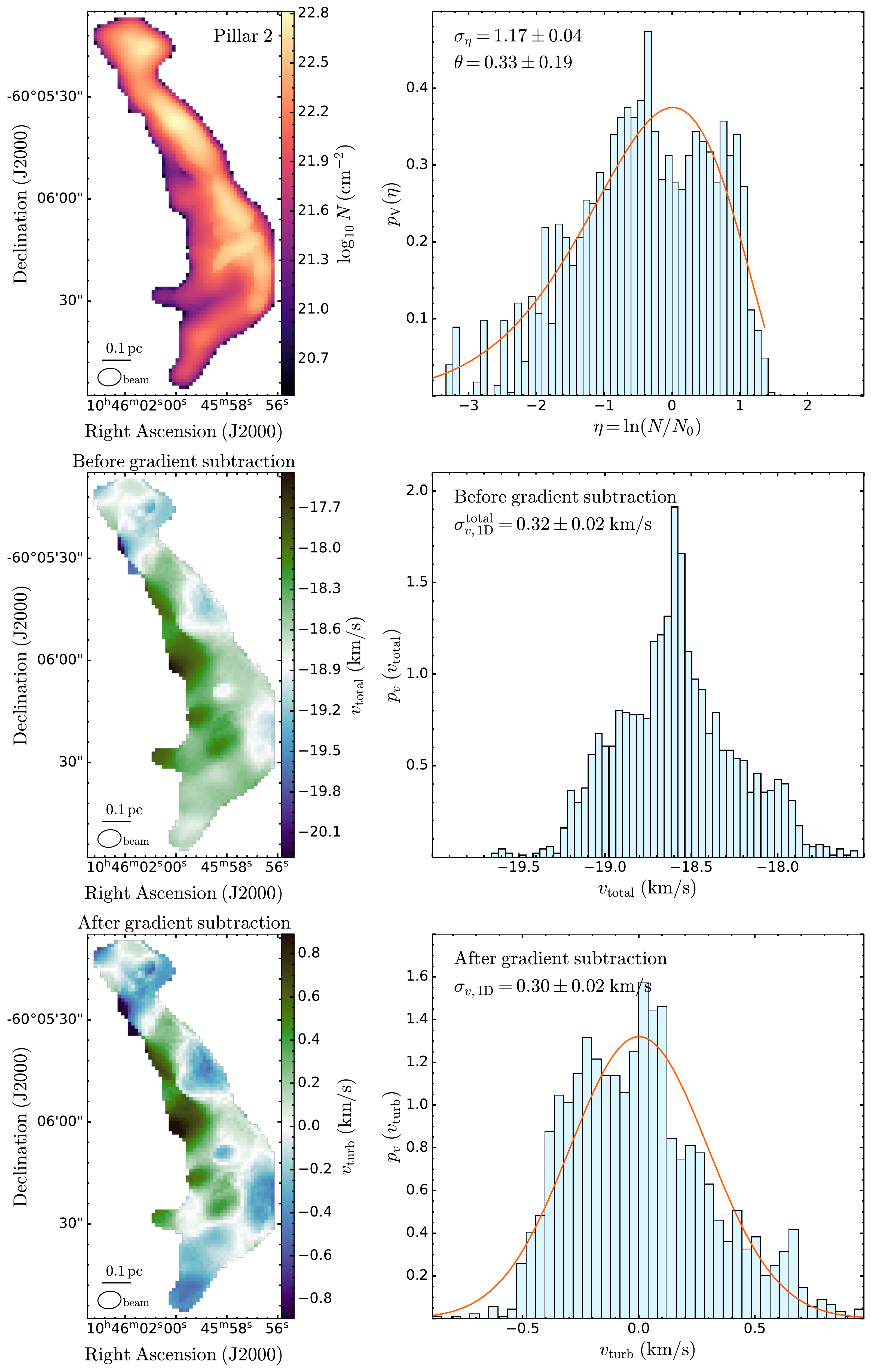}
    \caption{Same as Figure~\ref{fig:Pillar20_Maps}, but for pillar 2.}
    \label{fig:Pillar2_Maps}
\end{figure*}

The column density and first-moment maps for pillar~2 are shown in Figure~\ref{fig:Pillar2_Maps}. One feature that is interesting to note is that the $\mathrm{H}_2$ column density drops significantly as we move from the overdensity at the north-west to the pillar edge. This is because pillar~2 is irradiated by an O~star to the west of the image, which is photo-evaporating the $\mathrm{H}_2$ gas with its FUV radiation, leading to the sharp interface. In addition, relative to other pillars in the region, pillar~2 has a relatively high mean column density ($N_0 = 1.15 \pm 0.13 \, \mathrm{cm}^{-2}$). This is understandable from the column density map in Figure~\ref{fig:Pillar2_Maps}, wherein evident overdensities are visible that could/would be undergoing self-gravitational collapse. In fact, the column density PDF (scaled by $N_0$) shows an excess in the high-density part, and thus deviates from a PDF that is expected from pure supersonic turbulence. We thus speculate that pillar~2 is in a later evolutionary stage wherein self-gravity is dominating the local dynamics of the overdensities. This potentially leads to a much higher value of the density dispersion (see Table~\ref{tab:pillars_summary}), and hence a higher value of $b \sim 1.7$ than that expected from pure turbulence (i.e., \mbox{$b\sim 0.3$--$1.0$}). Ideally, to obtain the true purely turbulent density dispersion of pillar~2, it would be required to account for the overdense cores in the pillar, e.g., by fitting column-density profiles (for instance using dendograms). Those could then be subtracted out, similar to what we do for the velocity field. However, these procedures are outside the scope of this study, and instead, we see how much the effect of gravitational contraction can skew $b$ towards values greater than 1.

\subsection{Pillar 4}
\label{sec:Pillar4}

Pillar~4 is another member of the "Southern Pillars" and lies to the west of pillar~8 (discussed below), and as in most of the other pillars, shows signs of ongoing star formation in its vicinity. \citet{Smith_2010} found a bright rim of $\mathrm{H}\alpha$ emission from an associated HH object, HH~1008, and \citet{Ohlendorf_2012} identified the potential source to lie in a globule, which seems to be separating out from the bulk of the pillar. However, we do not include this globule in our analysis region, and instead pick only the bulk of the pillar. This is because the globule is disconnected from the bulk of the pillar, which compromises the accuracy of using our method. In addition, \citet{Klaassen_2020} report that the velocity dispersion of the $\mathrm{CO}$ at the location of the protostar is an order of magnitude greater than that of the rest of the pillar, suggesting a potential line of sight molecular outflow. Accounting for the outflow and separating out the turbulent motions in the globule simultaneously with the kinematically dissimilar bulk of the pillar is potentially very difficult. Instead, we focus only on the more interesting bulk of the pillar, which contains enough dense gas to potentially form stars in the future. 

\begin{figure*}
    \centering
    \includegraphics[width=0.8\textwidth]{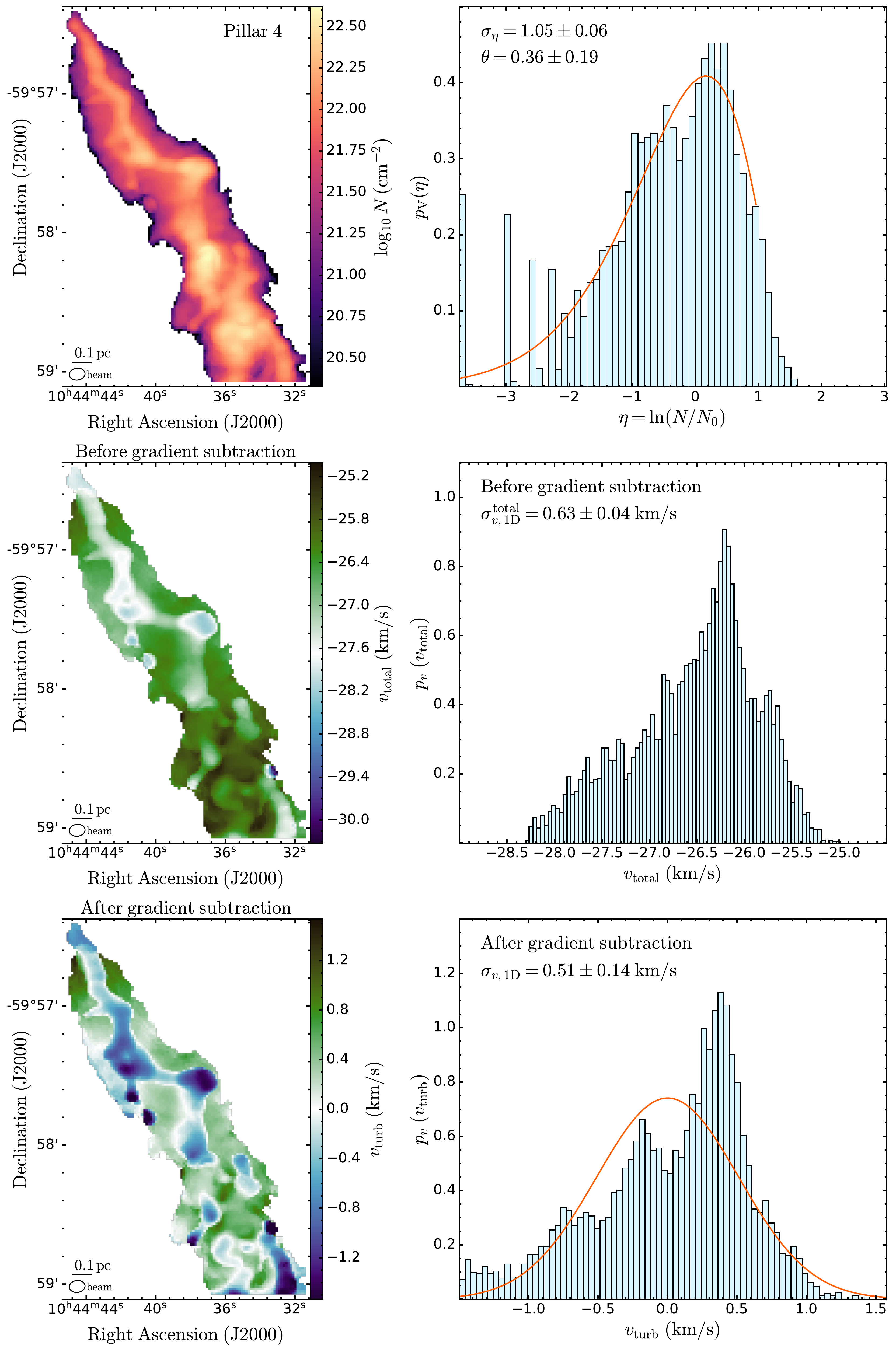}
    \caption{Same as Figure~\ref{fig:Pillar20_Maps}, but for pillar 4.}
    \label{fig:Pillar4_Maps}
\end{figure*}

In Figure~\ref{fig:Pillar4_Maps}, we show the column density and first-moment maps for the region that we study in pillar~4. Note that the globule containing the protostar lies outside the field-of-view of the figure in the north-east direction. We obtain a mean column density of $N_0 = 8.6 (+1.9/-1.8) \times 10^{21}\, \mathrm{cm}^{-2}$, and a corresponding total mass of $160 (+56/-46)\, \mathrm{M}_{\sun}$. Following the methods outlined in Section~\ref{sec:b_method}, we obtain a 3D volume density dispersion $\sigma_{\rho/\rho_0} = 3.1 (+2.1/-0.9)$ and a turbulent Mach number $\mathcal{M} = 3.9 (+1.0/-0.7)$. Combining these two quantities, we obtain $b = 0.8 (+0.6/-0.3)$. This suggests that the turbulence in pillar~4 is predominantly compressive in nature. We predict that the region in pillar~4 studied here is at an earlier evolutionary state than pillar~2, characterised by a lack of evident cores and higher values of $\alpha_\mathrm{vir}$ and $t_{\mathrm{ff}}/t_{\mathrm{turb}}$ (see Table~\ref{tab:pillars_summary}), suggesting that self-gravity is not yet dominating the local dynamics of the pillar. We speculate that the compressive turbulence in pillar~4, similar to pillar~22, is in the process of creating local overdensities that should collapse and lead to future sites of star formation, eventually leading to an evolutionary state as in pillar~2 wherein self-gravity dominates the local dynamics and density structure. 


\subsection{Pillar 8}
\label{sec:Pillar8}

Pillar~8 as we characterise in this study is part of the extensively studied dust pillar G287.84-0.82. It has a dense compact cluster embedded in its head, which is contributing to the local dynamic and thermodynamic state of the pillar \citep{Smith_2005,Hartigan_2015,Mookerjea_2019}. This cluster, "Treasure Chest" as coined by \citet{Smith_2005}, contains more than 69 young stars, including at least one O-star \citep[e.g.;][]{Hagele_2004}, illuminating its own compact \HII region seen in $\mathrm{H\alpha}$ and in typical photon-dominated region (PDR) line emissions \citep{Thackeray_1950,Smith_2005,Mookerjea_2019}. While this cluster influences the PDR and molecular gas dynamics of pillar~8, the overall structure is consistent with it being shaped by the feedback from $\eta\mathrm{Car}$ and the Trumpler~16 cluster. In fact, it has even been proposed that the Treasure Chest could itself have been triggered into formation due to the photoionising feedback of the aforementioned stars \citep{Smith_2005}. However, the high estimated age of $\sim 1.3 \, \mathrm{Myr}$ \citep{Oliveira_2018} for the Treasure Chest suggests that this is unlikely, and instead the cluster belongs to the same generation of star formation as $\eta\mathrm{Car}$ and Trumpler~16, albeit at an earlier evolutionary state. 

\begin{figure*}
    \centering
    \includegraphics[width=0.85\textwidth]{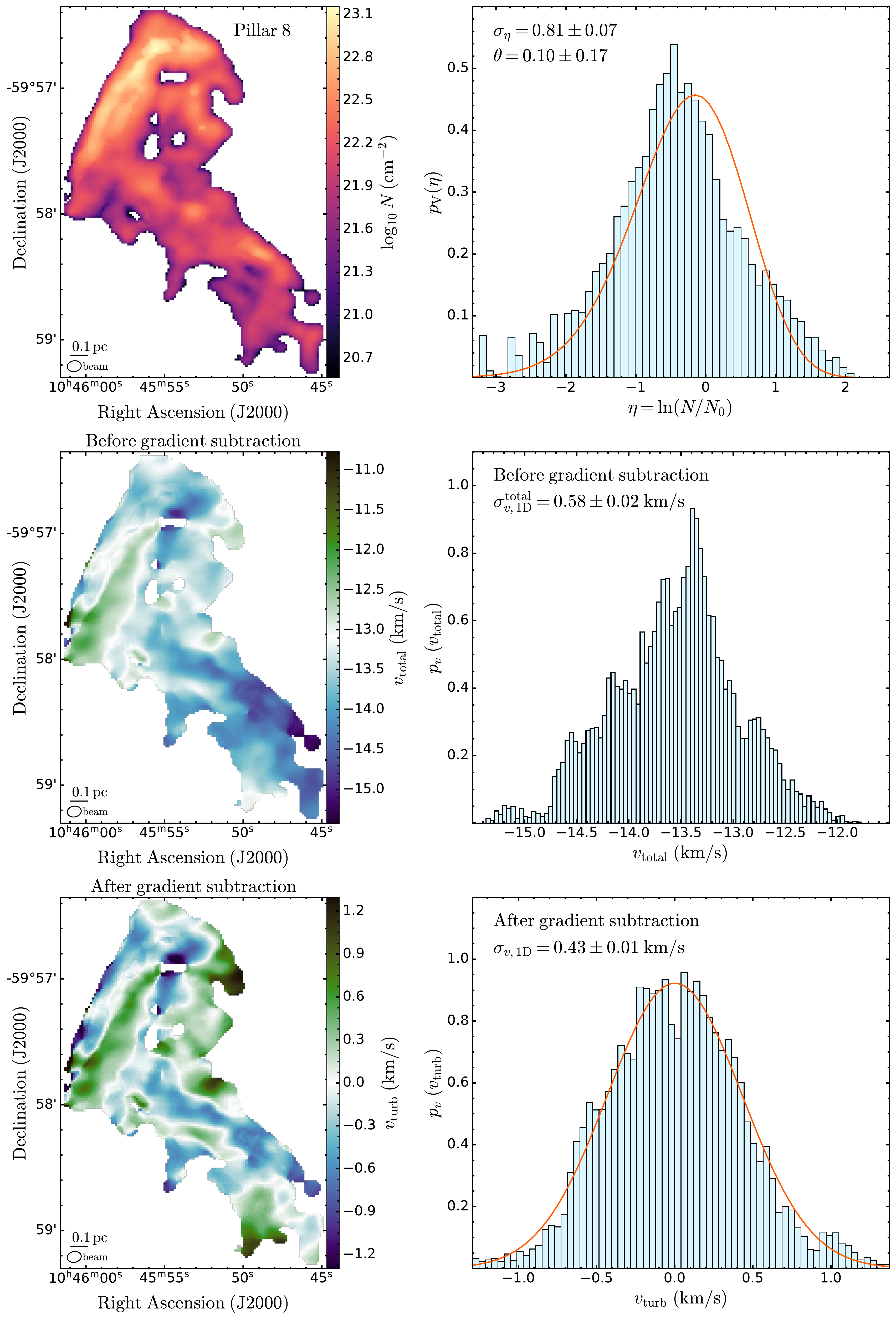}
    \caption{Same as Figure~\ref{fig:Pillar20_Maps}, but for pillar 8.}
    \label{fig:Pillar8_Maps}
\end{figure*}

That being said, the influence of the Treasure chest on the local gas column density and effectively on $b$ is evident. \citet{Mookerjea_2019} showed that a PDR is formed internally in pillar~8, primarily due to feedback from the Treasure Chest. Additionally, they observe a peak in dust emission and the high density line emissions traced with isotopologues of $\mathrm{CO}$, which are spatially slightly offset from the peak in PDR tracer emission. This morphologically arc-shaped overdensity in molecular gas that surrounds the estimated location of the internal PDR can be seen in our column density maps as well (see Figure~\ref{fig:Pillar8_Maps}). We find a mean column density for pillar~8 of $N_0 = 1.75 (+0.39/-0.38) \times 10^{22} \, \mathrm{cm}^{-2}$. This is in reasonable agreement with other studies that probed the column density structure in pillar~8 with molecular line emission \cite[e.g.;][]{Rathborne_2004,Rebolledo_2016} and dust continuum observations \citep[e.g.;][]{Roccatagliata_2013,Schneider_2015}. We find a total mass of $M = 511 (+182/-147) \mathrm{M}_{\sun}$ for our pillar~8 region, which is very close to the estimate in \citet{Mookerjea_2019}.

Our column density PDF fit yields a $\sigma_\eta = 0.81 (+0.07/-0.01)$ and $\theta = 0.10 \pm 0.17$, which corresponds to $\sigma_{N/N_0} = 0.79 (+0.13/-0.08)$. We compute $\mathcal{R} = 0.24 \pm 0.1$ from the density power spectrum and obtain $\sigma_{\rho/\rho_0} = 3.3 (+2.2/-1.0)$. This is combined with our Gaussian fit-based estimate of $\sigma_{v,3\mathrm{D}}= 0.75 (+0.02/-0.01) \, \mathrm{km}/\mathrm{s}$, and yields a Mach number $\mathcal{M} = 3.1 (+0.7/-0.4)$. The combination of $\sigma_{\rho/\rho_0}$ and $\mathcal{M}$ yields $b = 1.1 (+0.7/-0.4)$. Similar to pillar~2, the best-fit value of $b$ is slightly greater than $1$ for pillar~8, suggesting that there are density enhancements that are not just due to purely compressive turbulence, but that have contributions from gravitational collapse. Moreover, we find that the presence of the Treasure cluster embedded within pillar~8 potentially produces additional \textit{local} compressions through its UV feedback and/or winds, which might contribute to the density structure. These local non-turbulent compressions could lead to a wider density PDF, and this subsequently leads to a somewhat higher value of $b$ than cause by the external compression alone.

\subsection{Pillar 44}
\label{sec:Pillar44}
Pillar~44 is a small region that morphologically looks similar to one of the finger-like structures of pillar~22. It protrudes from a dust rim that lies on the south-western wall of Carina \citep{Hartigan_2015} and is facing in the general direction of Trumpler~14 and Trumpler~16. However, \citet{McLeod_2016} who studied this region with MUSE, suggests that the large projected distances to these clusters ($\sim 13 \, \mathrm{pc}$), and the fact that pillar~44 does not point towards their central coordinates, could indicate that the cluster is not the primary ionising source of pillar~44. They rather attribute the primary source to two other O-type stars: an O6 and and O9.5 star (HD303316 and HD305518), which are located $\sim 2.7 \, \mathrm{pc}$ to the north-east of the pillar (see figure~2 in their paper). In addition, pillar~44 is shown to host a protostar in its tip,  characterised by a highly-collimated protostellar jet (HH1010), as seen in the HST $\mathrm{H}\alpha$ image from \citet{Smith_2010}. The same jet was also observed with MUSE in \citet{McLeod_2016}, who estimate a dynamical age of $\sim 0.1 \, \mathrm{Myr}$ for the jet, and \citet{Reiter_2017} estimate an even younger age than this for HH1010. These estimates are significantly younger than the estimated ages for the central clusters, which could be an indication that the protostar was formed after the formation of the cluster.

Similar to pillar~22, the rim of the cloud from which pillar~44 is protruding, has a velocity offset from the pillar. However, unlike the case of pillar~22, it is not a smooth systematic gradient (suggestive of large-scale rotation/shear), but rather a systematically distinct velocity component. In addition, the base of the pillar is not completely interconnected when we apply our signal-to-noise cutoff, and hence we remove the base region completely from our analysis. 
This makes the net region to analyse the smallest among the regions in this work. However, although the region is small it is quite dense, with a mean column density $N_0 = 1.0 (+0.23/-0.22) \times 10^{22} \, \mathrm{cm}^{-2}$. Our fit to the PDF of $\eta$ yields $\sigma_\eta = 0.80 (+0.11/-0.17)$ and $\theta = 0.33 \pm 0.21$, which corresponds to $\sigma_{N/N_0} = 0.57 (+0.09/-0.04)$. Pillar~44 has a relatively high $\mathcal{R} = 0.35 \pm 0.14$, which leads to $\sigma_{\rho/\rho_0} = 1.7 (+1.1/-0.5)$. The unique feature of pillar~44 is its very low velocity dispersion $\sigma_{v,\mathrm{3\mathrm{D}}} = 0.32 (+0.02/-0.01) \, \mathrm{km}/\mathrm{s}$ and Mach number $\mathcal{M}=1.3 (+0.3/-0.2)$, significantly lower than for the other pillars. This low velocity dispersion leads to a value of $b\sim 1.2 (+0.8/-0.4)$, again slightly greater than 1. It is possible that this is due to the fact that pillar~44 is the densest and most evolved sub-region within a larger pillar that includes the rim. In addition, there are only \mbox{$\sim 5$--$6$} telescope beams across the length of this pillar, and it is possible that we are not resolving the kinematic and density structure of the gas sufficiently well, and so we must interpret the results for this pillar with some caution \citep[see][for a discussion on resolution requirements for our method]{Sharda_2018}.

\begin{figure*}
    \centering
    \includegraphics[width=0.9\textwidth]{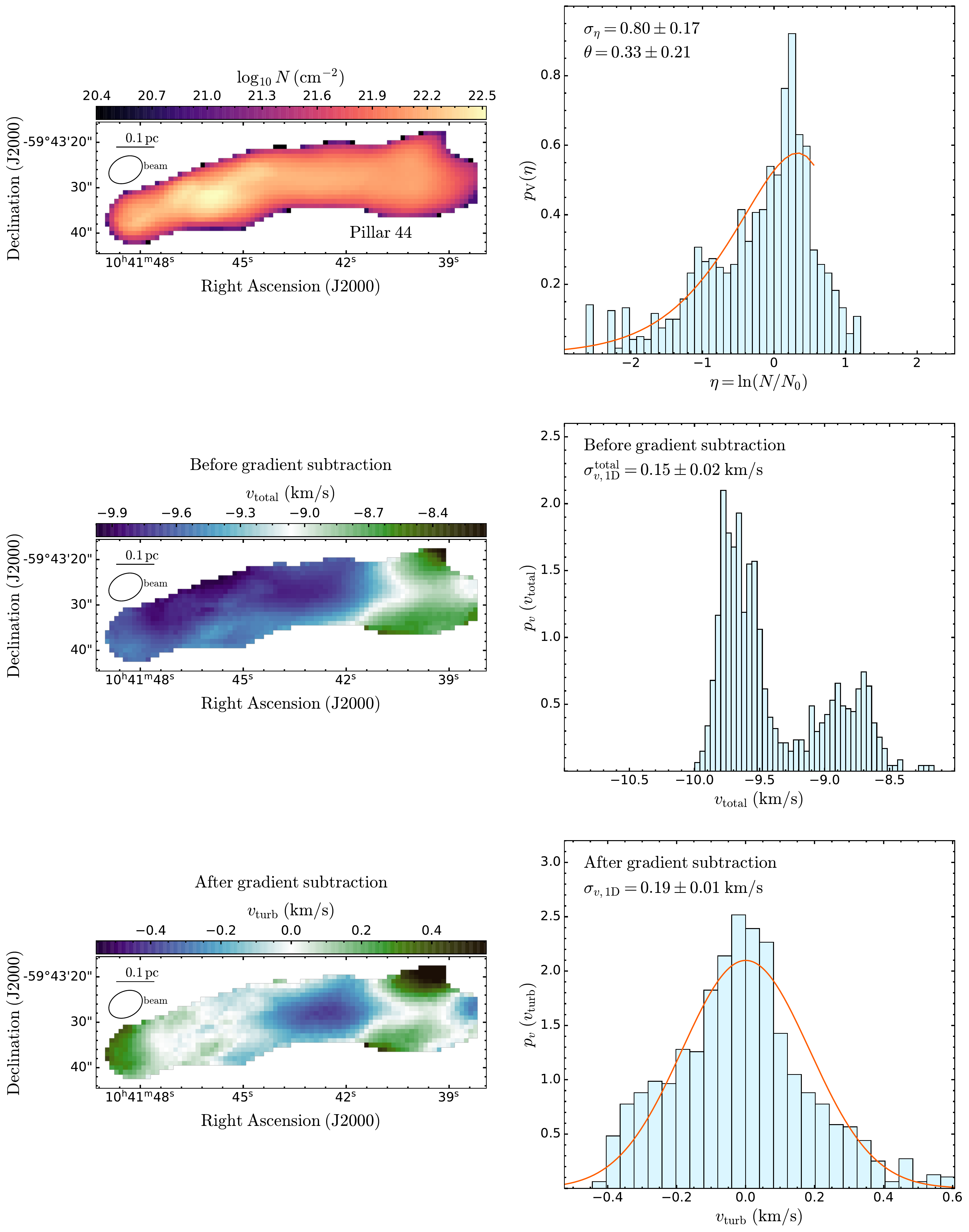}
   \caption{Same as Figure~\ref{fig:Pillar20_Maps}, but for pillar 44.}
    \label{fig:Pillar44_Maps}
\end{figure*}

\section{Discussion}
\label{sec:Discussion}
\subsection{Physical parameters of the pillars}
In Table~\ref{tab:pillars_summary} we summarise some key physical quantities of the pillar regions relevant to their dynamical states. We can see that the pillars span a range of masses (\mbox{$\sim 40$--$500 \, \mathrm{M}_{\sun}$} and effective areas (\mbox{$\sim 0.2$--$1.3 \, \mathrm{pc}^2$}). Note that the mass estimates in this study are lower than that reported in \citet{Klaassen_2020}, as we use a slightly different method to trace the $\mathrm{H}_2$ column density, with a higher column density threshold to focus on the denser regions of the pillars, and thus effectively probe a smaller region. We can see that pillar~2 and pillar~8 are comparatively denser than the other pillars. We attribute the high $N_0$ of these pillars to the high-column overdensities that are present in their structure, which seem to be largely gravitationally driven, suggesting that the pillars are at a later evolutionary stage than the other pillars, and/or there are significant non-turbulent contributions to their density structure. This is also evident from the low value of the virial parameter ($\alpha_{\mathrm{vir}}$) for pillar~2, indicating that it is gravitationally unstable. In the specific case of pillar~8, we suggest that the presence of a potential compact \HII region due to the O~stars present in the Treasure Chest cluster it shrouds \citep{Hagele_2004,Smith_2005,Mookerjea_2019}, strongly influence the dynamics of the pillar, creating an arc of high density, clearly visible in Figure~\ref{fig:Pillar8_Maps}. This again leads to a virially unstable state for pillar~8 ($\alpha_{\mathrm{vir}} \sim 0.6$). 

Another interesting quantity in Table~\ref{tab:pillars_summary} to highlight is the 3D velocity dispersion obtained directly from the mean value of the second-moment map $\sigma_{v,3\mathrm{D}}^{\mathrm{mom}2}$. We can see that this is higher than that we obtain after the plane-of-sky gradient/bulk-motion subtraction, followed by a subsequent Gaussian fit (see Section~\ref{sec:b_method_kinematics}), and we stress that we would be overestimating the turbulent velocity dispersion if we did not subtract the gradient in the first-moment map or if we used the second-moment map.

\subsection{The turbulence driving ($b$) and its role for the SFR}
\label{sec:b_SFR}

In this study we are primarily interested in the dominant mode of turbulence present in the pillars, quantified by the turbulence driving parameter $b$ given by Equation~\ref{eq:b_parameter}. The value of $b$ is an important quantity as the flow dynamics, density structure and the subsequent star formation rate (SFR) in a cloud depend on it, and is often used as a proxy for the efficiency of star formation in a cloud \citep[see][and references therein]{Federrath_2016}. We find median values of $b$ in the range $\sim 0.68 - 1.7$, summarised in Table~\ref{tab:pillars_summary}. This strongly suggests that the turbulence driving mode is primarily compressive (i.e., $b > 0.4$), and we can rule out mixed (i.e., $b\sim0.4$) and solenoidal (i.e., $b\lesssim1/3$) driving at the 1-sigma level. The viable agent proposed to drive this turbulence is the photoionising radiation of the massive stars in Carina through two potential physical mechanisms: at early times, the enhanced thermal pressure due to an expanding ionisation front compresses the gas in the pillars and drives turbulence; and alternatively, through the "rocket-effect" wherein ionising photons photo-evaporate gas at the edges of the pillars, which subsequently drives a shock into the pillar due to momentum conservation. Carina, however, has a highly porous shell that allows for significant leakage of ionised gas, quantified by a low value of the confinement factor \citep[see][]{Rosen_2014}. This suggests that the latter mechanism would be the more dominant one at the current dynamical state of Carina. That being said, the current turbulent density structure would have been shaped by the former mechanism as well at earlier times, and the current state could still have imprints of its influence. Thus, we do not comment on the exact physical mechanism of the two that drives this turbulence. Rather, we more generally attribute the driving source of turbulence that encompasses these two mechanisms to photoionising radiation. 

To demonstrate the significance of $b$ in the context of star formation, we calculate for pillar~22 an estimate for the star formation rate (SFR) per free-fall time $\epsilon_{\mathrm{ff}}$, i.e., the star formation rate normalised by the free-fall time of the gas, rendering the quantity dimensionless. In turbulence-regulated theories of star formation, $\epsilon_{\mathrm{ff}}$ is fully quantified by four dimensionless parameters: the virial parameter $\alpha_{\mathrm{vir}}$, the sonic Mach number $\mathcal{M}$, the turbulence driving parameter ($b$), and the plasma beta $\beta$ \citep{federrath_klessen_2012}. The value of $\epsilon_{\mathrm{ff}}$ with the multi-freefall model \citep{Hennebelle_Chabrier_2011} is given by equation~(41) in \citet{federrath_klessen_2012}, \begin{equation}
\epsilon_{\mathrm{ff}}=\frac{\epsilon}{2 \phi_{t}} \exp \left(\frac{3}{8} \sigma_{s}^{2}\right)\left[1+\operatorname{erf}\left(\frac{\sigma_{s}^{2}-s_{\mathrm{crit}}}{\sqrt{2 \sigma_{s}^{2}}}\right)\right],
\end{equation}
where $s_{\mathrm{crit}}$ is the log-normalized critical density threshold, $\sigma_s$ the log-normalized density variance, $\phi_t$ a parameter with value $1/\phi_t = 0.46 \pm 0.06$ \citep[from table~3 in][]{federrath_klessen_2012}, and $\epsilon=0.5$ the core-to-star efficiency \citep{Federrath_2014}. The critical density threshold is given by
\begin{equation}
s_{\mathrm{crit}}=\ln \left[\frac{\pi^{2}}{5} \phi_{x}^{2} \alpha_{\mathrm{vir}} \mathcal{M}^{2} \frac{1}{1+\beta^{-1}}\right],
\end{equation}
where $\phi_x$ is a parameter of order unity \citep{Krumholz_Mckee_2005,Padoan_2011}, fitted as $\phi_x = 0.17 \pm 0.02$ in \citet{federrath_klessen_2012}. The log-normalized density variance $\sigma_s$ is given by 
\begin{equation}
    \sigma_s^2 = \ln \left( 1 + b^2 \mathcal{M}^2 \beta/(\beta+1) \right).
\end{equation}
We use the values of $\mathcal{M} = 3.5 (+0.8/-0.5)$, $\alpha_{\mathrm{vir}} = 3.2 \pm 1.0$, and $b = 0.9 (+0.6/-0.3)$ for pillar~22 from Table~\ref{tab:pillars_summary} to calculate its $\epsilon_{\mathrm{ff}}$. We set the plasma beta $\beta = \infty$, as we do not have information on the magnetic fields in these pillars. With these values, we obtain $s_{\mathrm{crit}} = 0.7 \pm 0.2$, and a corresponding $\epsilon_{\mathrm{ff}} = 0.5 \pm 0.3$. This corresponds to an absolute $\mathrm{SFR} = \epsilon_\mathrm{ff}M/t_{\mathrm{ff}} = (3.5 \pm 1.8) \times 10^{-5} \, \mathrm{M}_{\sun}\, \mathrm{yr}^{-1}$ for pillar~22. We stress the reader to interpret this value with caution, and note that the purpose of this demonstration is not to provide an estimate of the current star formation rate of the pillars. Rather, the crucial point to note is that, if the turbulence mode was primarily a natural mixture of solenoidal and compressive modes (i.e., \mbox{$b\sim 0.4$}) rather than predominantly compressive, we would obtain an $\mathrm{SFR} = (1.4\pm 0.7) \times 10^{-5} \, \mathrm{M}_{\sun}\, \mathrm{yr}^{-1}$, which is almost a factor of 3 lower. This indicates that the driving of compressive turbulence in the pillars by photoionising radiation may enhance the efficiency with which they eventually form stars. 

 \subsection{Triggered or spontaneous star formation?}
 \label{sec:triggered}
 
While direct signatures of star formation through jets and outflows have been observed at the tip of pillars and/or globules in Carina, there is still the open question of whether this observed star formation is triggered or whether the stars spontaneously formed from the turbulent overdensities in the natal molecular cloud. While photoionisation feedback may seem like a viable agent to trigger the formation of stars, it is certainly plausible that the stars we observe in the pillars could have already been formed and the photoionising feedback just uncovers the envelope of dense gas that shrouds it. Thus, the question arises whether the mechanism of local triggering of star formation in the pillars is physically plausible, and if so, how it fits into the picture of the turbulent model of star formation in the ISM \citep[see][for a review]{Padoan_2014}. 

In this context, in \citetalias{Menon_20} we showed that a plane-parallel photoionisation front impinging onto a turbulent neutral gas cloud drives compressive modes of turbulence locally in the pillars, and raises the value of $b$ from the value that was present inherently in the natal molecular cloud. We suggested that since this value of $b$ is increased as a result of the turbulence driven by the photoionising radiation, indicates that the local dynamics of the region is altered in a way as to assist star formation, and thus this effectively triggers star formation in the pillars. Here we find that all the observed pillars are dominated by compressive modes of turbulence. We speculate that the pillars reached this turbulent state through the influence of the photoionising radiation, which raised the effective turbulence driving parameter $b$, as was seen in the simulations of \citetalias{Menon_20}. This further suggests that photoionising feedback could trigger local star formation. One may thus infer that stars form more effectively in the UV feedback-environments such as Carina, and indeed this has been claimed in some observational studies. For instance, \citet{Roccatagliata_2013} deduced from \textit{Herschel} wide-field maps that the star formation rate per unit mass in dense clouds in Carina is higher than that of other star-forming regions. Similarly, \citet{Xu_2019} find that the core formation efficiency (CFE) is systematically higher in UV-feedback dominated regions in M16, and is also on average higher than the CFE in GMCs not associated with strong UV feedback. 

While this is the viable explanation, it is possible that the inherent turbulence in Carina was already compressive, and we are just observing the imprint of the dynamics of the natal cloud. However, earlier studies of the value of $b$ of the dense gas in Milky Way spiral-arm clouds find values of \mbox{$b \sim 0.4$--$0.7$}, with the upper limit of the range found in star-forming regions and \HII regions \citep[e.g.][]{Padoan_1997,Brunt_2010_Taurus,Ginsburg_2013}, with younger molecular clouds estimated to have a mixed mode of turbulence, i.e., $b\sim 0.4$ \citep{Kainulainen_2013}. We note that these studies use the second-moment map to estimate the velocity dispersion rather than the first-moment map method that accounts for large-scale bulk motion. Thus, the values of $b$ in these studies are in a sense lower limits, as $\sigma_v$ would only be overestimated with the second-moment map in the presence of large-scale motions, and at best, remain the same in the absence of them. Simulations that study the formation and evolution of molecular clouds show that the value of $b$ corresponds to a mixture of solenoidal and compressive modes \citep{Pan_2016}. However, $b$ could increase within $\sim 5\, \mathrm{Myr}$ \citep{Kortgen_2017}. Interestingly, in a recent study, \citet{Orkisz_2017} showed that the large-scale turbulence in the Orion~B molecular cloud is inherently solenoidal or mixed, however, in the main star-forming regions NGC~2023 and NGC~2024, and the photo-dissociation front of the Horsehead nebula, the turbulence was shown to be strongly compressive, which could be a result of feedback-driven turbulence. Thus, while it is certainly possible that the inherent turbulence was compressive in Carina, we speculate that it is more plausible that the current turbulent dynamics is a result of the feedback from the massive stars in Carina. That being said, one can only speculate on the natal state of Carina, and hence similar studies of the turbulence driving mode in other potential/active star-forming environments are important to give a clearer picture of how feedback influences the turbulent dynamics of the gas.

\subsection{Evolutionary state of the pillars and the turbulence driving parameter}
It is interesting to see that although the values of $b$ of all the pillars in this study lie in the regime of compressive turbulence driving, there is a large range of $b$ values, with some pillars even slightly exceeding $b=1$. We speculate that this dynamic range exists, because the pillars studied in this work could be at different stages of their evolution. Some are relatively young and compressive turbulence driven by photoionising radiation is producing local overdensities efficiently, which could lead to future star formation (for instance in pillars~4, 20, and 22). However, some pillars are at later evolutionary stages wherein the formation of star-forming cores is well underway, and local gravitational collapse dominates the dynamics rather than compressive turbulence (for instance in pillar~2). This suggests that the value of $b$ rises locally in the pillar as it evolves (a feature also found in the simulations in \citetalias{Menon_20}). This happens until a certain point, after which local overdensities evolve in isolation, dominated by local gravitational collapse. We speculate that these overdensities are the precursors of the globules, evaporating gaseous globules and eventually proplyds that are observed in \HII regions \citep[see, for instance,][]{Schneider_2016}.
 

\subsection{Possible caveats}

\subsubsection{Magnetic fields}

Magnetic fields are seen ubiquitously in molecular clouds and play an important role in the structure and star formation of molecular clouds  \citep[see][for a comprehensive review]{Krumholz_Federrath_2019}. Ideally, we should be including a realistic value for the magnetic field strength to account for its effects. However, observations of the magnetic fields in the pillar structures are lacking, except for one study of the dense gas in the 'Pillars of Creation' in M16 \citep{Pattle_2018}. Hence we do not attempt to include the effects of magnetic fields in this work. Instead, we briefly summarise here their potential effects on photoionisation feedback and the driving of turbulence, and how our results could be influenced if we were to consider the contribution of magnetic fields.

It has been shown that the presence of magnetic fields could render the injection of kinetic energy to the cold molecular gas by the expansion of \HII regions more effective, subject to certain orientations of the magnetic field \citep{Gendelev_2012}. Although it has not yet been demonstrated that this results in stronger turbulence driving, the analogous process has been demonstrated for stellar wind feedback \citep{Offner_2018}. In addition, while it has been demonstrated that magnetic fields do not necessarily maintain turbulent motions on their own \citep{Stone_1998,MacLow_1998}, it is possible that they alter the decay rate of turbulence driven by the compression of gas that could be due to self-gravity or in our case radiative compression due to external pressure \citep{Birnboim_2018}. The physical mechanism is that, compression in a strongly magnetised flow causes the flow to become highly anisotropic \citep{Beattie_2019}, and this anisotropy could reduce the decay rate of supersonic turbulence. This could potentially have consequences for the star formation efficiency and lifetime of the pillars. Observational and numerical studies that include the effects of magnetic fields in the context of the formation and evolution of pillars are pertinent. We intend to investigate this in future works.  

Most importantly in the direct context of this study, the effective turbulence driving parameter $b$ in the presence of magnetic fields has an additional contribution due to magnetic pressure through the thermal-to-magnetic pressure ratio $\beta$, and is given by \citep{Padoan_2011,Molina_2012,federrath_klessen_2012}, 
\begin{equation}
\label{eq:b_magnetic}
    b = \frac{\sigma_{\rho/\rho_0}}{\mathcal{M}}(1+\beta^{-1})^{1/2},
\end{equation}
which reduces to Equation~\ref{eq:b_parameter} in the limit $\beta \to \infty$. It is evident that the inclusion of a realistic non-zero value for the magnetic field would only increase the value of $b$. \citet{Pattle_2018} suggest \mbox{$\beta \sim 1$--$10$} of the plane-of-sky magnetic fields in the pillars of M16 they observe. Using this range of values in Equation~\ref{eq:b_magnetic}, we can see that our value of effective $b$ would increase by \mbox{$\sim 4$--$40\%$}.

\subsubsection{Effect of assumed isotopologue abundances}
\label{sec:Abundances}
In this work we use $\mathrm{CO}$ and its isotopologue line emissions to trace the $\mathrm{H}_\mathrm{2}$ column density. However, this approach is prone to some difficulties, as column density PDFs obtained from molecular lines are shown to not be well confined, because they depend strongly on the choice of excitation temperature and abundances \citep[see][and references therein]{Schneider_2015}. Here we discuss the issues associated with this, and how our results could be affected by the assumed abundances (for a discussion of the excitation temperature, see Section~\ref{sec:Excitation_Temperature} below).

There are primarily two processes that can lead to local/global variations in the CO to $\mathrm{H}_2$ abundance ratio $X[\mathrm{H}_2/^{12}\mathrm{CO}]$. At high volume densities, CO is prone to depletion as it condenses on ice crystals in the ISM. At lower volume densities, processes such as the competition between $\mathrm{CO}$ formation and destruction, carbon isotope exchange and isotope selective photodissociation by FUV photons, makes it difficult to predict accurate values for the abundances. We would like to stress that the mean column densities we report for the pillars in Table~\ref{tab:pillars_summary} do not take into account these uncertainties in the elemental abundances, i.e., the uncertainties in $X[^{12}\mathrm{CO}/^{13}\mathrm{COs}]$,    $X[^{12}\mathrm{CO}/\mathrm{C}^{18}\mathrm{O}]$ and $X[\mathrm{H}_2/^{12}\mathrm{CO}]$. This underestimates the uncertainty in the column densities in our pillars and the mean column density $N_0$ we report, as the uncertainty in the abundances could be quite significant (\mbox{factor $\sim2$--$3$}), and would propogate linearly to our estimate of $N_0$. However, changes in the adopted abundances do not affect our value of $b$ and our interpretation of the turbulence driving mode, as the uncertainties in $N_0$ neither propagate to the final value of $b$ nor to its uncertainty. This is because the quantities involved in the calculation of $b$ are defined such that $N_0$ is divided out, so they merely quantify the column (or volume) density contrast, independent of $N_0$. 

However, spatial and/or environmental variations of the abundances within the pillars could be important in the context of $b$. For instance, \citet{Ripple_2013}
report that while $^{13}\mathrm{CO}$ provides a reliable tracer of $\mathrm{H}_2$ gas in the strongly self-shielded interior of clouds (with $A_v$ > 3 mag), it potentially misses out on a significant percentage (40--50\%) of $\mathrm{H}_2$ gas in the extended cloud envelope, an effect more pronounced in the presence of strong FUV radiation fields. This indicates that we could be underestimating the column density in the boundary surfaces of our pillars, as a result overestimating $\sigma_{N/N_0}$, and effectively overestimating the value of $b$. Alternatively, it is also possible that the high column density regions traced in our pillars are prone to depletion of CO on icy dust grains \citep[e.g.,][]{Caselli_1999}, or saturation effects that are enhanced due to the presence of edge-to-centre temperature gradients in the clouds \citep{Pineda2010}. It has been shown for instance, that $\mathrm{CO}$ underestimates the $\mathrm{H}_2$ column density at $A_v$ > 10 mag (corresponding to $\sim 10^{22}\, \mathrm{cm}^{-2}$) in the Taurus molecular cloud \citep{Pineda2010} and $A_v$>40 mag in Cygnus \citep{Schneider_2016}. If this holds in the case of Carina, it might indicate that we are underestimating the high column density tails in our PDFs, and as a result underestimating the value of $b$. 

While it is important to quantify the aforementioned effects, it is not straightforward and is outside the scope of this study. It is however important to keep in mind the caveats involved with using molecular line tracers in tracing the $\mathrm{H}_2$ column density in molecular clouds \citep{Goodman_2009,Schneider_2015}.


\subsubsection{Excitation temperature}
\label{sec:Excitation_Temperature}

Another caveat introduced by the use of molecular line emission to trace column densities is the uncertainty in the appropriate excitation temperature. In this study, we use the same value for the excitation temperatures \mbox{$T_\mathrm{ex} \sim 10$--$30$} for the $^{13}\mathrm{CO}$ and $\mathrm{C}^{18}\mathrm{O}$, based on the range of gas kinetic temperatures ($T_\mathrm{gas}$) in LTE, an assumption valid at higher column densities. Similar to the case of assuming abundances as discussed in Section~\ref{sec:Abundances}, a systematically different value of $T_\mathrm{ex}$ in the whole pillar would not affect our derived value of $b$. Instead, it is local environmental variations of $T_\mathrm{ex}$ that are relevant in the calculation of $b$. To investigate whether the choice of local $T_\mathrm{ex}$ alters our results, we recompute the values of $b$ by producing column density maps with $T_\mathrm{ex}$ obtained from the $^{12}\mathrm{CO}, \, J= 2-1$ emission main-beam temperature ($T_\mathrm{MB}$) assuming it is optically thick by \citep[equation~88 in][]{Mangum2015},
\begin{equation}
    T_\mathrm{ex} = \frac{h\nu/k}{\ln \left( 1 + \frac{h\nu/k}{T_\mathrm{MB} + J \left( T_\mathrm{BG} \right)} \right)} ,
    \label{eq:excitation_temperature}
\end{equation}
where $J \left( T_\mathrm{BG} \right)$ is obtained from Equation~\ref{eq:J_equation} for the background CMB radiation $T_\mathrm{BG} = 2.7 \, \mathrm{K}$. We find that our values of $b$ change by at most 10\% for some pillars, and are thus well within the 1-sigma uncertainties quoted in Table~\ref{tab:pillars_summary}. We can understand this weak dependence from the fact that any dependence on $T_\mathrm{ex}$ for the column density in a pixel is encapsulated in the pre-factor $f\left( T_\mathrm{ex} \right)$ given in Equation~\ref{eq:Prefactor}. However, $f\left( T_\mathrm{ex} \right)$ varies by at most $20\%$ in the range of excitation temperatures \mbox{$T_\mathrm{ex} \sim 10$--$70 \, \mathrm{K}$}, a range much wider than obtained from Equation~\ref{eq:excitation_temperature} for our pillars. This indicates that our conclusions are not significantly affected by the assumed value of $T_\mathrm{ex}$.

\section{Summary and conclusions}
\label{sec:Summary}
In this study, we use recent high-resolution ALMA observations ($6 \arcsec$ angular resolution, corresponding to $\sim 0.07 \, \mathrm{pc}$ for the distance of $\sim 2.5 \, \mathrm{kpc}$ to Carina) of the $^{12}\mathrm{CO}$, $^{13}\mathrm{CO}$, and $\mathrm{C}^{18}\mathrm{O}, \, J=2-1$ molecular emission lines, in six pillar-like structures at various spatial locations in the Carina Nebula, to probe the turbulent density and velocity structure of the pillars. We use a combination of the $^{13}\mathrm{CO}$ and $\mathrm{C}^{18}\mathrm{O}$ emission to trace the column density and intensity-weighted line-of-sight (LOS) velocities in the plane-of-sky (i.e., first-moment maps). We reconstruct the 3D density dispersion scaled by its mean $\sigma_{\rho/\rho_0}$ from the derived 2D column density maps using the column-density power spectrum and the method by \citet{Brunt_2010b}. We then compute the 3D turbulent velocity dispersion in the molecular gas $\sigma_{v,3\mathrm{D}}$, accounting for, and subtracting contributions from non-turbulent motions in the first-moment map. Using these, we calculate the turbulence driving parameter ($b$) using Equation~\ref{eq:b_parameter}. Our main results can be summarized as follows:
\begin{enumerate}
    \item We observe pillars of varying sizes and masses in our study, with a range of mean column densities $N_0$  \mbox{$\sim 3$--$10 \times 10^{21}\mathrm{cm}^{-2}$} and supersonic turbulent Mach numbers \mbox{$\mathcal{M} \sim 1.3$--$4$} (see Table~\ref{tab:pillars_summary}). 
    \item We obtain a range of median values for $b$ in the pillars from \mbox{$b \sim 0.7$--$1.7$}, and a minimum value of $b\sim0.5$, considering the 1-sigma uncertainties (see Table~\ref{tab:pillars_summary}). This suggests that photoionising radiation drives predominantly compressive modes of turbulence in the cold, dense molecular gas of the pillars in Carina.  
    \item Values of $b>1$ for some pillars indicate that there are significant non-turbulent contributions to the high-density part of the PDF, which we attribute to local self-gravitational collapse. We speculate that these higher values are obtained for pillars in later stages of their dynamical evolution, wherein the gas overdensities that were seeded by compressive turbulence locally start to collapse, and self-gravity dominates their evolution later on. This is evident in our study, wherein more gravitationally bound pillars (as characterised by their virial parameter $\alpha_\mathrm{vir}$ and/or ratio of freefall time to turbulent crossing time) have systematically higher values of $b$.
\end{enumerate}
Our main observational finding that predominantly compressive turbulence is driven by photoioinising radiation in the pillars is in line with similar conclusions from numerical simulations \citepalias{Menon_20}. The fact that the dynamical mechanisms of photoionisation feedback increase the fraction of compressive modes in the turbulence strengthens the interpretation that the active and potential sites of \textit{local} star formation observed in these pillars have been triggered by the photoionising radiation. Follow-up studies of the turbulent gas dynamics and its relation to the star formation efficiency of feedback-dominated regions would allow for more quantitative predictions on the net effect of photoionization on star formation in molecular clouds, and its potential to trigger \textit{local} star formation. 

\section*{Acknowledgements}
We thank the anonymous referee for a very constructive review that improved the quality of this paper. SHM would like to thank James Beattie for insightful discussions on the manuscript. C.~F.~acknowledges funding provided by the Australian Research Council (Discovery Project DP170100603 and Future Fellowship FT180100495), and the Australia-Germany Joint Research Cooperation Scheme (UA-DAAD).
RK acknowledges financial support via the Emmy Noether Research Group on Accretion Flows and Feedback in Realistic Models of Massive Star Formation funded by the German Research Foundation (DFG) under grant no. KU 2849/3-1 and KU 2849/3-2.
This paper makes use of the following ALMA data: ADS/JAO.ALMA\#2016.1.00101.S. ALMA is a partnership of ESO (representing its member states), NSF (USA) and NINS (Japan), together with NRC (Canada) and NSC and ASIAA (Taiwan), in cooperation with the Republic of Chile. The Joint ALMA Observatory is operated by ESO, AUI/NRAO and NAOJ. This research made use of \verb|Astropy|,\footnote{\url{http://www.astropy.org}} a community-developed core \verb|Python| package for Astronomy \citep{astropy_2013,astropy_2018}. 

\section*{Data availability}
The data underlying this article will be shared on reasonable request to the corresponding author.



\bibliographystyle{mnras}
\bibliography{pillar_driving} 




\appendix


\section{Dependence on the signal-to-noise cutoff}
\label{sec:appendix_SNcutoff}
\begin{figure*}
    \centering
    \includegraphics[width=0.7\textwidth]{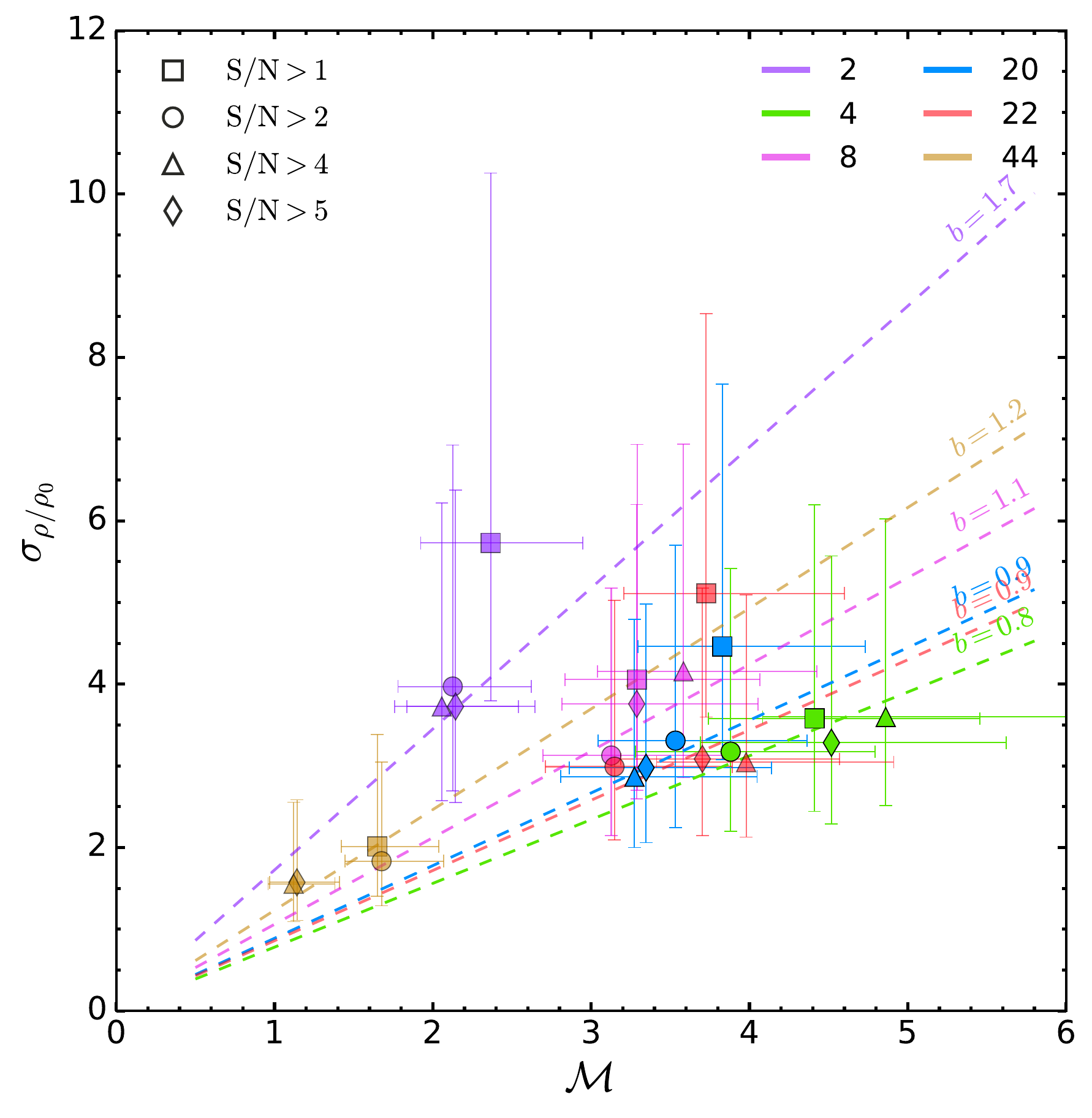}
    \caption{Comparison of the values of the 3D density dispersion $\sigma_{\rho/\rho_0}$ and the turbulent Mach number $\mathcal{M}$ obtained for the pillars by adopting different signal-to-noise cutoffs, or equivalently different column density thresholds. We show the values for $\mathrm{S/N}>1$ (squares), $\mathrm{S/N}>2$ (circles), $\mathrm{S/N}>4$ (triangles), $\mathrm{S/N}>5$ (diamonds) for the different pillars (see colours in the legend). For comparison we plot representative lines of Eq.~(\ref{eq:b_parameter}) for each pillar, corresponding to the median values of $b$ obtained with our fiducial $\mathrm{S/N}>3$ cutoff. We can see that, although there is slight scatter from the fiducial median value for the different cases, overall the value of $b$ is well constrained within the 1-sigma error bars.}
    \label{fig:SN_Cutoff}
\end{figure*}

Here we discuss the dependence of our results on the adopted signal-to-noise (S/N) cutoff, or equivalently on the $\mathrm{H}_2$ column density threshold. In the study we use a fiducial value of S/N of 3, which corresponds to a column density threshold of $\sim 1.1 \times 10^{20}\, \mathrm{cm}^{-2}$. However we also explore higher and lower values of S/N, of 1, 2, 4, and 5, which correspond to column density thresholds of $\sim 0.37, 0.74, 1.5, \, \mathrm{and} \, 1.9 \times 10^{20} \, \mathrm{cm}^{-2}$, respectively. In Figure~\ref{fig:SN_Cutoff} we show the values of $\sigma_{\rho/\rho_0}$ and $\mathcal{M}$ obtained for the different cases of S/N, with representative lines of $b$ corresponding to the fiducial values for the different pillars quoted in Table~\ref{tab:pillars_summary}. We find that, although there is slight scatter from the fiducial median value for the different cases, overall the value of $b$ is well constrained within the 1-sigma error bars.

\section{Molecular line constants}
\label{sec:appendix_line_constants}
Here we list the line constants for the calculation of the column density from the line emission fluxes discussed in Section \ref{sec:b_method_density_structure}.  The main beam brightness temperature $T_\mathrm{MB}$ is converted to a $\mathrm{CO}$ isotopologue column density using Equation~(\ref{eq:Column_Density}) using a pre-factor $f\left(T_\mathrm{ex} \right)$ (see Equation~\ref{eq:Prefactor}), where $T_{\mathrm{ex}}$ is the excitation temperature. For the $^{13}\mathrm{CO} \; 2-1$ transition we use $\nu = 220.4 \; \mathrm{GHz}$, $\mu = 0.112 \; \mathrm{Debye}$, and an upper level energy $E_\mathrm{up}/k_{\mathrm{B}} = 16.6 \, \mathrm{K}$, corresponding to the $^{13} \, \mathrm{CO} \; J=2$ rotational energy level. Similarly for the $\mathrm{C}^{18}\mathrm{O} \; 2-1$ transition we use $\nu = 219.6 \; \mathrm{GHz}$, $\mu = 0.1098 \; \mathrm{Debye}$, and an upper level energy $E_\mathrm{up}/k_{\mathrm{B}} = 15.8 \, \mathrm{K}$, corresponding to the $\mathrm{C}^{18}\mathrm{O} \; J=2$ rotational energy level. We use $T_\mathrm{BG} = 2.7 \, \mathrm{K}$ to correct for the CMB background emission.  


\bsp	
\label{lastpage}
\end{document}